\documentclass[12pt]{article}
\usepackage{unformatted}
\usepackage{url}
\date{September 25, 2020}
\usepackage{hyperref}
\hypersetup{
    colorlinks=true,
    linkcolor=blue,
    filecolor=magenta,
    urlcolor=cyan,
    citecolor=blue,
}
\usepackage[numbers]{natbib}
\usepackage{a4}
\usepackage{ifthen}
\renewcommand{\cite}[2][nobrackets]{\ifthenelse{\equal{#1}{nobrackets}}{\Citet{#2}}{\Citep{#2}}}

\newcommand{\articletype}[1]{}
\newcommand{\received}[1]{}
\newcommand{\revised}[1]{}
\newcommand{\accepted}[1]{}
\newcommand{\corres}[1]{}

\usepackage{amssymb}

\usepackage{amsmath}

\usepackage{color}

\usepackage{tikz}
\usepackage{pgfplots}

\newlength{\figureWidth}
\newlength{\figureHeight}

\usepackage[pagewise]{lineno}

\newcommand*\patchAmsMathEnvironmentForLineno[1]{%
  \expandafter\let\csname old#1\expandafter\endcsname\csname #1\endcsname
  \expandafter\let\csname oldend#1\expandafter\endcsname\csname end#1\endcsname
  \renewenvironment{#1}%
     {\linenomath\csname old#1\endcsname}%
     {\csname oldend#1\endcsname\endlinenomath}}%
\newcommand*\patchBothAmsMathEnvironmentsForLineno[1]{%
  \patchAmsMathEnvironmentForLineno{#1}%
  \patchAmsMathEnvironmentForLineno{#1*}}%
\AtBeginDocument{%
\patchBothAmsMathEnvironmentsForLineno{equation}%
\patchBothAmsMathEnvironmentsForLineno{align}%
\patchBothAmsMathEnvironmentsForLineno{flalign}%
\patchBothAmsMathEnvironmentsForLineno{alignat}%
\patchBothAmsMathEnvironmentsForLineno{gather}%
\patchBothAmsMathEnvironmentsForLineno{multline}%
}

\journal{Journal of Computational Physics}

\usepackage{hyperref}

\newcommand{\brx}[1]{\left(#1\right)}
\newcommand{\pd}[2]{\frac{\partial #1}{\partial #2}}

\newcommand{\Exp}[1]{e^{#1}}

\def\R{\mathbb{R}}
\def\E{\mathbb{E}}

\DeclareMathOperator*{\orth}{orth}

\newcommand\bvec[1]{\textbf{#1}}
\newcommand\sbvec[1]{\bar{\bvec{#1}}}
\newcommand\tbvec[1]{\tilde{\bvec{#1}}}

\renewcommand{\eqref}[1]{(\ref{eq:#1})}
\newcommand{\figref}[1]{Figure \ref{fig:#1}}
\newcommand{\tabref}[1]{Table \ref{tab:#1}}

\newcommand{\secref}[1]{Section \ref{sec:#1}}

\definecolor{dgreen}{rgb}{0,0.7,0}
\newcommand{\llabel}[1]{}\newcommand{\refonecol}[1]{\textcolor{black}{#1}}
\newcommand{\reftwocol}[1]{\textcolor{black}{#1}}
\newcommand{\refthreecol}[1]{\textcolor{black}{#1}}
\newcommand{\refone}[2]{\llabel{#1}\refonecol{#2}}
\newcommand{\reftwo}[2]{\llabel{#1}\reftwocol{#2}}
\newcommand{\refthree}[2]{\llabel{#1}\refthreecol{#2}}

\usepackage{subfig}
\let\subtop\subfloat
\usepackage{graphbox}

\usepackage{framed}
\usepackage{caption}
\usepackage[plain]{algorithm}
\usepackage[noend]{algpseudocode}
\usepackage{color,colortbl}
\definecolor{dblue}{rgb}{0,0,0.7}

\usepackage{ifthen}
\renewcommand{\cite}[2][nobrackets]{\ifthenelse{\equal{#1}{nobrackets}}{\Citet{#2}}{\Citep{#2}}}

\begin{document}

\begin{frontmatter}

\title{Application of Adaptive Multilevel Splitting to High-Dimensional Dynamical Systems}

\author[rug]{S.~Baars}\ead{s.baars@rug.nl}
\author[uu]{D.~Castellana}\ead{d.castellana@uu.nl}
\author[rug]{F.W.~Wubs}\ead{f.w.wubs@rug.nl}
\author[uu,ccss]{H.A.~Dijkstra}\ead{h.a.dijkstra@uu.nl}

\address[rug]{Bernoulli Institute for Mathematics, Computer Science and Artificial Intelligence,
  University of Groningen, P.O. Box 407, 9700 AK Groningen, The Netherlands}

\address[uu]{Institute for Marine and Atmospheric research Utrecht,
  Department of Physics,  Utrecht University,
  Princetonplein 5, 3584 CC Utrecht, The Netherlands}

\address[ccss]{Centre for Complex Systems Studies, Utrecht University, Leuvenlaan 4
3584 CE Utrecht, The Netherlands}

\begin{abstract}
  Stochastic nonlinear dynamical systems can undergo rapid transitions relative to the change in their forcing, for example due to the occurrence of multiple equilibrium solutions for a specific interval of parameters.
  In this paper, we modify one of the methods developed to compute probabilities of such transitions, Trajectory-Adaptive Multilevel Sampling (TAMS), to be able to apply it to high-dimensional systems.
  The key innovation is a projected time-stepping approach, which leads to a strong reduction in computational costs, in particular memory usage.
  The performance of this new implementation of TAMS is studied through an example of the collapse of the Atlantic Ocean Circulation.
\end{abstract}


\end{frontmatter}

\section{Introduction}
The Atlantic Meridional Overturning Circulation (AMOC) is an important component of the climate system because of its associated meridional heat transport.
\refthree{moc}{It is well known that the AMOC is sensitive to surface freshwater flux perturbations \cite[]{stommel:61}.
Freshening of the surface waters in the Nordic and Labrador Seas diminishes the production of deepwater that feeds the deep southward branch of the AMOC.
The weakening of the AMOC leads to reduced northward salt transport freshening the northern North Atlantic and amplifying the original freshwater perturbation.
In a hierarchy of ocean models, transitions in AMOC flows are found due to the occurrence of multiple equilibria (see e.g.\ chapter 6 in \cite{dijkstra:05}).
However, it is currently unknown whether the present-day AMOC is in a multiple equilibrium regime, and if so, what the probability is that it will undergo a collapse within the next 100 or 1000 years \cite[]{castellana:19}.}

Probabilities of rare events, and in particular of transitions between different equilibrium states of a certain stochastic system, are often very difficult to compute.
In the framework of Large Deviation Theory, there are several results available \cite[]{freidlin:84}.
When the deterministic system can be described in terms of a potential, the transition rate for stochastic transitions induced by white additive noise is given by the Eyring--Kramers formula \cite[]{eyring:35,kramers:40}.
Extensions of this formula for non-gradient systems have been found \cite[]{bouchet:16}, \refone{tone}{and although these can be very useful for determining the location of possible transition paths \cite[]{bouchet:19}, directly computing transition probabilities using such formulas for high-dimensional systems is presently not feasible.}
Furthermore, these results allow to compute only transition rates, i.e.\ the probability that a transition occurs in a unit time.
A more interesting quantity is the probability that the system undergoes a transition within a certain time interval.
\refone{exp}{The connection between these two quantities is not straightforward, unless specific assumptions are made on the noise generating the transitions: for example, if the process can be assumed to be Poisson, then the probability of undergoing a transition within a certain time follows the cumulative distribution function of the exponential distribution with the rate parameter being the transition rate \cite[]{lestang:18}.}

From a totally different perspective, one could think of computing transition probabilities with standard Monte Carlo methods.
This can be done by computing a large number of trajectories, and then counting the number of observed transitions.
This approach is, however, not suitable for high-dimensional systems, especially when the probabilities are small.
Therefore, specialized methods, such as Genealogical Particle Analysis (GPA) \cite[]{moral:05,moral:13,wouters:16}, Adaptive Multilevel Splitting (AMS) \cite[]{cerou:07, rolland:15,rolland:15b}, and Trajectory-Adaptive Multilevel Sampling (TAMS) \cite[]{lestang:18} have been developed.
The idea behind these methods is similar: the algorithms try to push trajectories in the direction of the transition to make sure transitions actually occur, also for smaller sample sizes, while still being able to compute the transition probability.
\refone{baro}{Rare transitions in a barotropic $\beta$-plane model have been studied by means of AMS in \cite[]{bouchet:19}.}
Transitions in the two-dimensional barotropic quasi-geostrophic equations were also studied in \cite{bouchet:14, laurie:15}, but in that case only the most likely transition path was determined by means of the minimum action method \cite[]{e:04,vanden-eijnden:08,zhou:08,grafke:17}.
To our knowledge, actual transition probabilities have never been computed for high-dimensional problems on the scale \refone{dof}{of our target problem, which is a global ocean model with a resolution of at least 2 degrees ($\sim 1.4\cdot 10^6$ degrees of freedom).
In this paper, however, we only present and validate the methodology suitable for the target problem, but do not actually compute transition probabilities for this problem.}

Methods for finding transition probabilities have two main \refone{limitations}{limitations}.
The first is that they always require some form of time integration, which is very expensive for high-dimensional systems.
This is especially the case when an implicit time-stepping method is used, since then in every time step a linear system has to be solved.
The other is that for methods based on AMS, a large number of states at different time steps have to be stored in memory \refone{tone2}{in order to keep the computational cost minimal.}
We therefore propose a projected time-stepping method, for which in every time step, only a small projected linear system has to be solved, and for which only the projected states have to be stored, which are much smaller dimensional than the original states.
\refone{modred1}{The vectors that are used for the projection are the empirical orthogonal functions (EOFs) that are commonly used often easily computed for many geophysical models.
The most dominant EOFs can be found by computing the eigenvectors belonging to the largest eigenvalues of the covariance matrices at both stable steady states.}
These vectors can be obtained from basis vectors of the low-rank solution of a generalized Lyapunov equation \cite{baars:17}.
\refone{papers1}{A theoretical background is given in \cite{hartmann:11}.}

The projected time-stepping method is related to the Karhunen--Lo\`eve transform in the context of stochastic processes \cite[]{loeve:55} or the proper orthogonal decomposition in the context of fluid dynamics \cite[]{lumley:67, cazemier:98}.
The difference is that we do not use an ensemble, but instead use both stable steady states and covariance matrices
at those steady states to directly compute the basis vectors.

\refone{papers2}{Model order reduction techniques such as the projection method described above have been applied in many fields when studying transitions \cite[]{hartmann:05, yvinec:12, mukhin:15, hartmann:16, mohamad:16, hernandez:18}, but to our knowledge not with the multilevel splitting methods that we need to compute transition rates in non-gradient systems.}
\refone{modred2}{Note that there are many other available model reduction techniques that could be used with the methodology described in this paper such as temporal EOFs \cite[]{mukhin:15} and machine learning \cite[]{hernandez:18}.}

In this paper, TAMS, its optimization and the projected time-stepping are described in section 2.
This is followed by section 3 where we apply it to a spatially two-dimensional model of the AMOC, governed by a set of stochastic partial differential equations.
Summary and discussion are provided in section 4.
\section{Methodology}

Consider a system of stochastic differential algebraic equations (SDAEs) of the form
\begin{align}\label{eq:SPDE}
M(\bvec{p}) \text{d} \bvec X_t = F(\bvec X_t; \bvec{p}) \text{d}t + g( \bvec X_t; \bvec{p}) \text{d} \bvec W_t
\end{align}
where $\bvec{p} \in \R^{n_p}$ indicates a vector of parameters, $M \in \R^{n \times n}$ is usually referred to as the mass matrix, $\bvec X_t \in \mathbb{R}^n$, $F: \mathbb{R}^n \to \mathbb{R}^n$, $g \in \mathbb{R}^{n \times n_w}$ and $\bvec W_t \in \mathbb{R}^{n_w}$ \refone{brownian}{is a vector of $n_w$ independent standard Brownian motions \cite[]{gardiner:85}.}
Assume that the corresponding deterministic system
\refthree{det}{
\begin{align}\label{eq:PDE}
  M(\bvec{p}) \frac{\text{d} \bvec x}{\text{d}t} = F(\bvec x; \bvec{p})
\end{align}
  }
has two stable steady states $\sbvec x_A$ and $\sbvec x_B$ for fixed values of parameters $\bvec{p}$ and define two sets, $A$ and $B$, which respectively contain $\sbvec x_A$ and $\sbvec x_B$.
Our starting point will be the state $\sbvec x_A$.
Once we perturb it with noise, the state will wander around the equilibrium, with a motion depending on the characteristics of the noise.
If the perturbation is strong enough, the system may undergo a transition and end up in the other equilibrium state $\sbvec x_B$.
We would like to compute the probability that such a transition occurs within a certain interval of time.

\refthree{continuation}{We compute bifurcation diagrams for \eqref{PDE} using the pseudo-arclength continuation method \cite[]{keller:77}, in which branches of steady states $(\bvec{x}(s), \bvec{p}(s))$ are parameterized by an arclength parameter $s$ and Newton's method is used on an augmented system to converge onto the steady states.
Pseudo-arclength continuation has the advantage that it can also compute the unstable steady state $\sbvec x_C$ that is located between $\sbvec x_A$ and $\sbvec x_B$.}

\subsection{Trajectory-Adaptive Multilevel Sampling}

The idea behind the algorithm is based on the fact that, once the trajectories leave $A$, there are more trajectories that come back to $A$ than there are ones that reach $B$.
A method that makes use of this is called the Adaptive Multilevel Splitting method (AMS) \cite[]{cerou:07, rolland:15}.
It was inspired by multilevel splitting methods which date back to \cite{kahn:51} and \cite{rosenbluth:55}.
The multilevel splitting methods are all based on the same idea of discarding trajectories that do not reach $B$ and splitting (or branching) trajectories that are closer \refthree{to}{to} $B$.
This makes it so that the probability that a trajectory reaches $B$ keeps increasing, which is why this method is more efficient than a naive method (such as Monte Carlo).

How close a vector $\bvec{x}$ is to $B$ is defined by a so-called reaction coordinate, or score function, which is a smooth one-dimensional function
\begin{align}
  \phi~:~\R^n\to\R.
\end{align}
In this paper, we assume that the reaction coordinate should satisfy
\begin{subequations}
  \begin{align}
    |\nabla\phi(\bvec{x})|&\neq 0,~\forall x\in\R^n\backslash(A\cup B),\\
    A&\subset \{\bvec{x}\in\R^n~:~\phi(\bvec{x})<z_{\min}\},\\
    B&\subset \{\bvec{x}\in\R^n~:~\phi(\bvec{x})>z_{\max}\},
  \end{align}
\end{subequations}
where $z_{\min}<z_{\max}$ are two given real numbers.
These properties were slightly adapted from \cite{cerou:11}.
For multi-dimensional problems, we propose some additional properties to make sure the method actually converges towards $B$, and not somewhere completely different which has the same value of $\phi$.
These properties are
\begin{subequations}
  \begin{align}
    \{\bvec{x}\in\R^n~:~\phi(\bvec{x})&=\inf\{\phi(\bvec{y})~:~\bvec{y}\in\R^n\}\}\subset A,\\
    \{\bvec{x}\in\R^n~:~\phi(\bvec{x})&=\sup\{\phi(\bvec{y})~:~\bvec{y}\in\R^n\}\}\subset B.
  \end{align}
\end{subequations}
Since the gradient $\nabla\phi$ is not allowed to be zero outside of $A$ and $B$, this means that the reaction coordinate is always increasing towards $B$ and decreasing towards $A$.

The efficiency of multilevel splitting methods is based on the choice of the reaction coordinate.
With $\alpha$ indicating the transition probability, \refone{err}{$N$ the number of trajectories and $\mu_{\alpha}$, $\sigma_{\alpha}$ the sample mean and sample variance of the estimator of $\alpha$, the relative error $\epsilon_\alpha=\sigma_{\alpha}/\mu_{\alpha}$} of standard Monte Carlo methods converges with $\mathcal{O}(\sqrt{1/(\alpha N)})$ \cite[]{rubino:09}, whereas multilevel splitting methods show behavior between $\mathcal{O}(\sqrt{\log{(1/\alpha)}/N})$ (optimal) and $\mathcal{O}(\sqrt{1/(\alpha N)})$ (worst case) \cite[]{simonnet:14, lestang:18}.
Choosing a better reaction coordinate means that we get closer to the optimal convergence behavior.
\refthree{committor}{The optimal behavior, however, is only attained when the optimal reaction coordinate, which is also referred to as the committor, is used, in which case the proportionality constant is equal to one \cite[]{rolland:15}.
The committor can be obtained by solving a backward Fokker--Planck equation, but in practice this can never be done for high-dimensional systems \cite[]{metzner:06}.}

In this paper, we define our reaction coordinate to be
\begin{align}
  \phi(\bvec{x}) = \eta - \eta \Exp{-\gamma d_A^2} + (1-\eta) \Exp{-\gamma d_B^2},\label{eq:react}
\end{align}
where $d_A=\Vert \bvec{x} - \sbvec{x}_A\Vert_2 / \Vert \sbvec{x}_A - \sbvec{x}_B\Vert_2$ is the normalized distance between $\bvec{x}$ and $\sbvec{x}_A$, and $d_B=\Vert \bvec{x} - \sbvec{x}_B\Vert_2 / \Vert \sbvec{x}_A - \sbvec{x}_B\Vert_2$ is the normalized distance between $\bvec{x}$ and $\sbvec{x}_B$ and $\gamma$ is a real positive constant which we choose to be 8.
\refthree{eta}{To give equal weight to both stable steady states, a good choice for $\eta$ would be 0.5.
When the unstable state $\sbvec{x}_C$ associated with the saddle-node bifurcation is known, however, a good choice would be $\eta=\Vert \sbvec{x}_C - \sbvec{x}_A\Vert_2 / \Vert \sbvec{x}_B - \sbvec{x}_A\Vert_2$.
This is the normalized distance between the unstable steady state $\sbvec{x}_C$ and the stable steady state $\sbvec{x}_A$.
Using this value of $\eta$ makes sure that more weight is given to the stable steady state farthest away from the unstable steady state, meaning that a trajectory is less likely to keep circling around it.}

AMS can be used to first compute the mean first passage time.
The transition probability can then be obtained from the mean first passage time under the assumptions that the transitions are a Poisson process and that they are instantaneous \cite[]{lestang:18}.

\begin{figure}[ht]
  \centering
  \includegraphics[width=\textwidth]{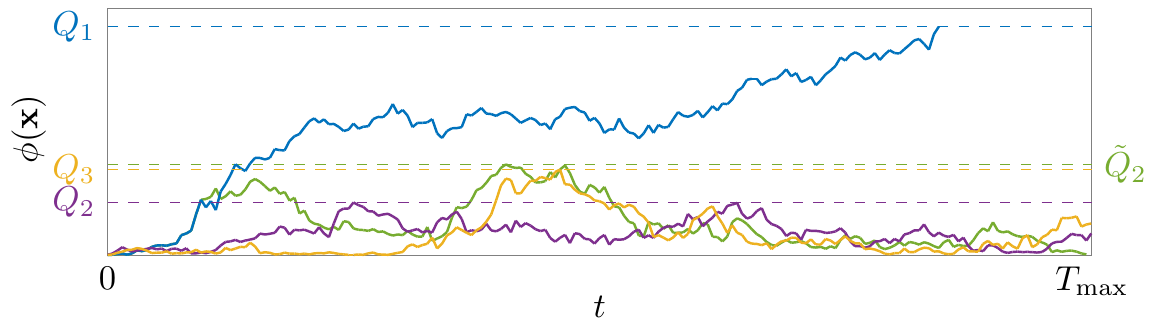}
  \caption{
    \refthreecol{
    Example run of TAMS with $N=3$ applied to a two-dimensional double well potential, for which we take $F(\bvec x) = -\nabla V(\bvec x)$ with $V(x,y) = \frac14 x^4-\frac12 x^2+y^2$ and $g(\bvec x)=0.4$.}
    The second trajectory (purple) has the lowest maximum value of the reaction coordinate $Q_2$ after the first iteration.
    This trajectory is then branched from a random other trajectory, in this case the first trajectory (blue).
    The new trajectory (green) has a new maximum value of the reaction coordinate $\tilde Q_2$, which is larger than $Q_2$.
    \label{fig:tams}}
\end{figure}

Trajectory-Adaptive Multilevel Sampling (TAMS) \cite[]{lestang:18} is a slightly altered version of AMS, in which, instead of first computing the mean first passage time, the transition probability is computed directly when given a maximum time $T_{\max}$, without requiring any of the aforementioned assumptions \cite{baars:19}.
In TAMS, there is an optional maximum number of iterations $k_{\max}$ that can be set, which allows us to stop even before all trajectories have reached $B$. We now explain step by step how to compute the transition probability using TAMS.
\begin{enumerate}
\item First generate $N$ independent trajectories $(\bvec{x})^{(1)},\ldots, (\bvec{x})^{(N)}$ with time step $\Delta t$ that start in $A$ until $t=T_{\max}$ is reached, or the trajectory ends up in $B$.
  Here $(\bvec{x})^{(i)} = (\bvec{x}_0^{(i)},\bvec{x}_1^{(i)},\ldots,\bvec{x}_{N_t}^{(i)})$ is a trajectory of length $N_t =T_{\max}/\Delta t$.
\item Each trajectory $(\bvec{x})^{(i)}$, $i=1,\ldots,N$, has a certain maximum value of the reaction coordinate (or maximum distance from $A$), which we define to be $Q_i$.
  Set $k=1$, $w_0=1$.
\item Take $L=\{j~:~Q_j=\inf\{Q_i~:~i\in \{1,\ldots,N\}\}\}$, which are the indices for which the maximum value of the reaction coordinate is minimal, and take $\ell_k$ to be the number of elements in $L$.
  Since the trajectories $\{(\bvec{x})^{(j)}~:~j\in L\}$ have the smallest value of $Q_i$, all other trajectories have some $j$ for which $\phi(\bvec{x}_j) \geq Q_{l}$ for all $l\in L$. \label{enum:tams-1}
\item Set $\displaystyle w_{k} = \left(1-\frac{\ell_k}{N}\right)w_{k-1}$.
  For all $l\in L$, repeat steps \ref{enum:tams-2}-\ref{enum:tams-3}.
\item Select a trajectory $(\bvec{x})^{(r)}$ uniformly at random with $r$ from $\{1,\ldots,N\} \backslash L$, and set $(\tbvec{x})^{(l)}=(\bvec{x}_0^{(r)},\ldots,\bvec{x}_{j_{\min}}^{(r)})$, where $j_{\min}$ is the smallest value for which $\phi(\bvec{x}_{j_{\min}}^{(r)})\geq Q_l$.\label{enum:tams-2}
\item Generate the rest of the trajectory starting from $\bvec{x}_{j_{\min}}^{(r)}$ until again you reach either $t=T_{\max}$ or $B$.
  This trajectory has a new maximum value of the reaction coordinate $\tilde Q_l$, which is always greater than or equal to $Q_l$.
\item Set $(\bvec{x})^{(l)}=(\tbvec{x})^{(l)}$ and $Q_l=\tilde Q_l$.\label{enum:tams-3}
  \label{enum:tams-4}
\item Repeat steps \ref{enum:tams-1}-\ref{enum:tams-4} with $k=k+1$ until $Q_i\geq z_{\max},~\forall~i=1,\ldots,N$ or $k=k_{\max}$.
\end{enumerate}
An illustration of the workings of TAMS is given in \figref{tams}.

The weights $w_k$, computed in each step, represent the probability of a trajectory reaching iteration $k+1$.
For instance, consider the case in which we start with 100 trajectories, and we have 2 trajectories for which the maximum value of the reaction coordinate is minimal.
Since these two trajectories are eliminated, the probability of a trajectory reaching iteration 2 is $1-2/100 = 98/100$.
We repeat this process, multiplying the probabilities that we find in every step.
This gives us an unbiased estimator of the transition probability
\begin{align}
  \hat\alpha_N=\frac{N_B w_{K}}{N} = \frac{N_B}{N}\prod_{k=0}^K\left(1-\frac{\ell_k}{N}\right),
\end{align}
where $K$ is the number of iterations it took for all trajectories to converge, and $N_B$ is the number of trajectories that reached $B$ \cite[]{lestang:18}.

\subsection{Optimizations}\label{sec:optim}
Say we have a problem of dimension $n = 10^4$ with a time step $\Delta t = 10^{-2}$, a maximum time $T = 10^3$, and $N = 10^3$ trajectories, which is not at all unreasonable for the problems that we want to solve.
In this case we would need at least 7.28 TB of memory to store all of the states that we compute \refthree{double}{in double precision}, which is a very large amount of memory.
Fortunately, some simple optimizations exist to help with this.

First, it is important to observe that the only place where we actually use a state is when we branch a trajectory.
Other than in this place, we only use the times to compute the quantities that we need.
This means that we can discard any state $\bvec{x}$ for which $\phi(\bvec{x})<Q_l$, since these will never be used for branching.
Discarding these states can be done for instance every time a trajectory is branched, every so many iterations, or even in every TAMS iteration.

Secondly, since we only use the first $\bvec{x}$ for which $\phi(\bvec{x})\geq Q_l$, we only have to store $\{\bvec{x}_j \in (\bvec{x}) ~:~ \phi(\bvec{x}_j) > \sup\{\phi(\bvec{x}_1),\ldots,\phi(\bvec{x}_{j-1})\}\}$, where $(\bvec{x})$ denotes a full trajectory.
This means that if we iteratively determine the value of $Q$, we only store a state $\bvec{x}$ if $\phi(\bvec{x})>Q$.
In addition, one could also limit the number of states that we store by only storing one state per interval of $\phi$.
So for instance if $z_{\min}=0.1$, $z_{\max}=0.9$ and we only store a state with intervals of 0.005, at most 160 states would be stored.

\refone{checkpointing}{Alternatively, one can implement a checkpointing strategy, where only a small subset of the states is stored, but all seeds of the random number generator, along with the corresponding values of $\phi(\bvec{x})$ are stored.
In that way, any state that is required for the branching process can be recomputed without actually storing it in memory.
In practice it is common to store the state that is being branched from, meaning that only one state per trajectory has to be stored.
Of course this comes at the cost of having to recompute $\bvec{x}_{j_{\min}}$, starting from the checkpoint, in every iteration of TAMS.
The storage in memory can be compressed even further by retaining only one checkpoint per group of trajectories that have a common ancestor, again at a slightly higher computational cost.}

Optimizations can also be done in terms of parallelization.
Since all trajectories in the first step are computed independently, parallelization of the first step is trivial.
Later steps, however, are more difficult to parallelize due to the branching process.
An example of a parallelized AMS method is given in \cite{rolland:15}.

\subsection{Accuracy of the method}\label{sec:error}
Generally, one realization of the algorithm is not enough to get an accurate answer.
Also, from just one realization, we can not tell how accurate the answer actually is.
Therefore, one computes $M$ realizations of the algorithm, and uses those to compute both a more accurate answer, and an estimate of the error.
From this we get a sample of $M$ probability estimates $\hat\alpha_N$, which we can use to compute the sample mean
\begin{align}
  \mu_{\alpha}=\frac{1}{M}\sum_{m=1}^M\hat\alpha_{N,m}
\end{align}
and the sample variance
\begin{align}
  \sigma_{\alpha}^2=\frac{1}{M-1}\sum_{m=1}^M\left(\hat\alpha_{N,m}-\mu_{\alpha}\right)^2.
\end{align}
An estimate of the relative error is given by
\begin{align}
  \epsilon_\alpha=\frac{\sigma_{\alpha}}{\mu_{\alpha}},
\end{align}
We noted before that the relative error shows behavior between $\mathcal{O}(\sqrt{\log{(1/\alpha)}/N})$ (optimal) and $\mathcal{O}(\sqrt{1/(\alpha N)})$ (worst case).

\refone{var}{It is known that when the committor is used as a reaction coordinate, the variance on the estimation of $\alpha$ tends to $\sqrt{|\log{\alpha}|}/\sqrt{N}$ for large $N$ \cite[]{cerou:07, rolland:15}.
This means that we can compute a compensated variance
\begin{align}
  \sigma_0=\sigma_{\alpha} \frac{\sqrt{N}}{\mu_\alpha\sqrt{|\log{\mu_\alpha}|}}
\end{align}
which can also be seen as the relative error divided by its expected optimal behavior.
Since behavior is optimal when the committor is used, in that case $\sigma_0$ would be equal to 1.
This quantity can be used as an indicator of the quality of the reaction coordinate.}

\subsection{Projected time-stepping in TAMS}
In AMS and TAMS, but also in GPA, one is free to choose any stochastic time stepping method.
\refone{theta}{
One such method is the stochastic theta method with $\theta\neq 0$ \cite[]{kloeden:92}, which, although convergence of the error of the rare event algorithm generally goes with $\sqrt{\Delta t}$, may be of benefit when large time steps are used \cite[]{rolland:15}.}
In case $\theta\neq 0$, multiple linear systems have to be solved in every time step.
Solving these linear systems is usually the computationally most expensive part of TAMS, even if the linear system is sparse, and an iterative solver with a good preconditioner is used.

At time step $j+1$ of the stochastic theta method performed on \eqref{SPDE} with fixed $\bvec{p}$, we take
\begin{align}\label{eq:stm}
  \hat F(\bvec{x}) = M\bvec{x}_j-M\bvec{x} + (1-\theta)\Delta t F(\bvec{x}_{j}) + \theta\Delta t F(\bvec{x}) + g(\bvec{x}_j)~\Delta\bvec{W}_{j}=0
\end{align}
with Jacobian matrix
\begin{align}
  \hat J(\bvec{x})=\theta\Delta t J(\bvec{x}) - M.
\end{align}
where $J$ is the Jacobian matrix of $F$ and $M$ is the mass matrix.
\refone{newton}{Newton's method} for one implicit time step would normally look like this
\begin{algorithm}[H]
  \centering
  \parbox{.95\textwidth}{
    \begin{framed}
      \begin{algorithmic}[1]
        \State $\bvec{y}^{(0)} = \bvec{x}_j$
        \For{$k=1,2,\ldots$ until convergence}
        \State \textbf{solve} $\hat J(\bvec{y}^{(k-1)}) \Delta \bvec{y}^{(k)} = -\hat F(\bvec{y}^{(k-1)})$
        \State $\bvec{y}^{(k)} = \bvec{y}^{(k-1)} + \Delta \bvec{y}^{(k)}$
        \EndFor
        \State $\bvec{x}_{j+1}=\bvec{y}^{(k)}$
      \end{algorithmic}
    \end{framed}}
\end{algorithm}
A small potential optimization that can be made is computing $\hat J(\bvec{x}_j)$ beforehand, and using this instead of $\hat J(\bvec{y}^{(k-1)})$ (the chord method).
This will make it such that Newton's method does not converge quadratically, but saves the time of computing the Jacobian matrix multiple times.

Instead of solving with this large sparse matrix $\hat J(\bvec{y}^{(k-1)})$ in every Newton iteration, we instead propose to solve with a smaller matrix obtained from a Galerkin type projection $V^T\hat J(\bvec{x}_j)V$, where $V$ consists of an orthogonal basis of the most important directions that are used by the stochastic perturbation.
\reftwo{lyap}{These vectors are given by the empirical orthogonal functions (EOFs), or more precisely by the eigenvectors belonging to the largest eigenvalues of the covariance matrix.
After linearization around the steady state $\sbvec{x}_A$, the covariance matrix $C_A$ may be obtained by solving a generalized Lyapunov equation of the form
\begin{align}\label{eq:gle}
  J(\sbvec{x}_A)C_AM^T+MC_AJ(\sbvec{x}_A)^T+B_AB_A^T=0,
\end{align}
where $B_A \in \R^{n\times n_w}$ is the matrix describing the noise \cite[]{gardiner:85}.
Instead of computing the full covariance matrix, a low-rank approximation of the form $C_A\approx V_AY_AV_A^T$, which only contains the eigenvectors belonging to the largest eigenvalues, may be computed instead \cite[]{baars:17}.}
\refthree{dims}{Here $V_A\in \R^{n \times n_A}$ and $Y_A\in \R^{n_A \times n_A}$ with $n_A \ll n$.}
The cost of computing such a low-rank approximation of the covariance matrix is negligible compared to the cost of computing transition probabilities.

\refone{basis1}{
We can now compute an orthogonal basis $V$ of the form
\begin{align}
  V = \orth{([\sbvec{x}_A, V_A, B_A, \sbvec{x}_B, V_B, B_B])},
\end{align}
where $\sbvec{x}_A$ and $\sbvec{x}_B$ are the two stable steady states, $V_A$ is the basis of the low-rank Lyapunov solution at $\sbvec{x}_A$, and $V_B$ is the basis of the low-rank Lyapunov solution at $\sbvec{x}_B$, and $B_A$ and $B_B$ are the noise matrices at $\sbvec{x}_A$ and $\sbvec{x}_B$.}
\refone{basis3}{The basis may be expanded further with information from around the saddle-node bifurcation if available.
Note that we can not compute $V_C$ in the same way as $V_A$ and $V_B$ because the steady state $\sbvec{x}_C$ is unstable.}

We can now replace the Newton iteration with
\begin{algorithm}[H]
  \centering
  \parbox{.95\textwidth}{
    \begin{framed}
      \begin{algorithmic}[1]
        \State $\bvec{y}^{(0)} = V^T\bvec{x}_j$
        \State $A=V^T\hat J(\bvec{x}_j)V$
        \For{$k=1,2,\ldots$ until convergence}
        \State \textbf{solve} $A \Delta \bvec{y}^{(k)} = -V^T\hat F(V \bvec{y}^{(k-1)})$
        \State $\bvec{y}^{(k)} = \bvec{y}^{(k-1)} + \Delta \bvec{y}^{(k)}$
        \EndFor
        \State $\bvec{x}_{j+1}=V\bvec{y}^{(k)}$
      \end{algorithmic}
    \end{framed}}
\end{algorithm}
If we apply this in TAMS, we see that between two consecutive time steps, we always apply $V^TV\bvec{y}^{(k)}$.
Now since $V$ is orthogonal, this means that we might as well apply TAMS itself to $\bvec{y}^{(k)}$ directly.
Doing this actually solves the most significant problem we observe when applying TAMS to high-dimensional systems, which is the storage of the trajectories, since we now only have to store the trajectories restricted to the space \refone{basis2}{spanned by} $V$.
Say we have a system of size 10000 and $V$ consisting of 100 vectors, which is a realistic scenario, then this reduces the required storage by a factor 100.

\section{Application to AMOC model}\label{sec:ocean}
In order to study the sensitivity of the AMOC to freshwater anomalies, we use the spatially quasi two-dimensional model as described in \cite{toom:11}.
In the model, there are two active tracers: temperature $T$ and salinity $S$, which are related to the density $\rho$ by a linear equation of state
\begin{align}
  \rho =\rho_0\brx{1-\alpha_T\brx{T -T_0}+\alpha_S\brx{S -S_0}},
\end{align}
where $\alpha_T$ and $\alpha_S$ are the thermal expansion and haline contraction coefficients, respectively, and $\rho_0$, $T_0$, and $S_0$ are reference quantities.

\subsection{Model formulation}

In order to eliminate longitudinal dependence from the problem, we consider a purely buoyancy-driven flow on a non-rotating Earth \cite[]{toom:11}.
We furthermore assume that inertia can be neglected in the meridional momentum equation.
The mixing of momentum and tracers due to eddies is parameterized by simple anisotropic diffusion.
In this case, the zonal velocity as well as the longitudinal derivatives are zero and the equations for the meridional velocity $v$, vertical velocity $w$, pressure $p$, and the tracers $T$ and $S$ are given by
\begin{subequations}
  \begin{align}\label{eq:GE}
    -\frac{1}{\rho_0 r_0}\pd{p}{\theta}+A_V\pd{^2v}{z^2}+\frac{A_H}{r_0^2}\brx{\frac{1}{\cos\theta}\pd{}{\theta}\brx{\cos\theta\pd{v}{\theta}}+\brx{1-\tan^2\theta}v}&=0,\\
    -\frac{1}{\rho_0}\pd{p}{z}+g\brx{\alpha_T T-\alpha_S S}&=0,\\
    \frac{1}{r_0\cos\theta}\pd{(v\cos\theta)}{\theta}+\pd{w}{z}&=0,\\
    \pd{T}{t}+\frac{v}{r_0}\pd{T}{\theta}+w\pd{T}{z}=\frac{K_H}{r_0^2\cos\theta}\pd{}{\theta}\brx{\cos\theta\pd{T}{\theta}}+K_V\pd{^2T}{z^2}&+\mathrm{CA}(T), \\
    \pd{S}{t}+\frac{v}{r_0}\pd{S}{\theta}+w\pd{S}{z}=\frac{K_H}{r_0^2\cos\theta}\pd{}{\theta}\brx{\cos\theta\pd{S}{\theta}}+K_V\pd{^2S}{z^2}&+\mathrm{CA}(S).
  \end{align}
\end{subequations}
Here $t$ is time, $\theta$ latitude, $z$ the vertical coordinate, $r_0$ the radius of Earth, $g$ the acceleration due to gravity, $A_H$ ($A_V$) the horizontal (vertical) eddy viscosity, and $K_H$ ($K_V$) the horizontal (vertical) eddy diffusivity.
The terms with CA represent convective adjustment contributions.

The equations are solved on an equatorially symmetric, flat-bottomed domain.
The basin is bounded by latitudes $\theta = - \theta_N$ and $\theta=\theta_N = 60^\circ$ and has depth $D$.
Free-slip conditions apply at the lateral walls and at the bottom.
Rigid lid conditions are assumed at the surface and atmospheric pressure is neglected.
The wind stress is zero everywhere, and ``mixed'' boundary conditions apply for temperature and salinity,
\begin{subequations}
  \begin{align}
    K_V\pd{T}{z}&=\frac{H_m}{\tau}\brx{\bar{T}(\theta)-T} , \\
    K_V\pd{S}{z}&= S_0 F_s(\theta).
  \end{align}
\end{subequations}
This formulation implies that temperatures in the upper model layer (of depth $H_m$) are relaxed to a prescribed temperature profile $\bar{T}$ at a rate $\tau^{-1}$, while salinity is forced by a net freshwater flux $F_s$, which is converted to an equivalent virtual salinity flux by multiplication with $S_0$.
The specification of the CA terms is given in \cite{toom:11}.
The numerical values of the fixed model parameters are summarized in \tabref{modelpars}.
\begin{table*}[tph]
  \centering
  \begin{tabular}{|rcll|rcll|}
    \hline
    $D   $ &=& $4.0   \cdot 10^3$    & m                & $H_m$      &=& $2.5   \cdot 10^2$    & m              \\
    $r_0 $ &=& $6.371 \cdot 10^6$    & m                & $T_0$      &=& $15.0$                & $^\circ$C      \\
    $g   $ &=& $9.8$                 & m s$^{-2}$       & $S_0$      &=& $35.0$                & psu            \\
    $A_H $ &=& $2.2   \cdot 10^{12}$ & m$^2$s$^{-1}$    & $\alpha_T$ &=& $1.0   \cdot 10^{-4}$ & K$^{-1}$       \\
    $A_V $ &=& $1.0   \cdot 10^{-3}$ & m$^2$s$^{-1}$    & $\alpha_S$ &=& $7.6   \cdot 10^{-4}$ & psu$^{-1}$     \\
    $K_H $ &=& $1.0   \cdot 10^3$    & m$^2$s$^{-1}$    & $\rho_0$   &=& $1.0   \cdot 10^3$    & kg m$^{-3}$    \\
    $K_V $ &=& $1.0   \cdot 10^{-4}$ & m$^2$s$^{-1}$    & $\tau$     &=& $75.0$                & days
    \\\hline
  \end{tabular}
  \caption{Fixed model parameters of the two-dimensional ocean model, following \cite{mheen:13}.}
  \label{tab:modelpars}
\end{table*}

\subsection{Bifurcation diagram} \label{sec:moc-bifdiag}
For the deterministic model, we take the surface forcing similar to what was used in \cite{mheen:13} as
\begin{subequations}
  \begin{align}
    \bar{T}(\theta) &= 10.0\cos\brx{\pi\theta/\theta_N},\\
    F_s(\theta) = {\bar F}_s(\theta) &=  \mu F_0 \frac{\cos(\pi\theta/\theta_N)}{\cos(\theta)} + \beta F_0 F_p(\theta),
  \end{align}
  \label{eq:moc-FSd}
\end{subequations}
where $\mu=0.3$ is the strength of the mean freshwater forcing, $F_0 = 0.1~\textrm{m yr}^{-1}$ is a reference freshwater flux, $\beta$ is the strength of an anomalous freshwater flux pattern $F_p$, which is \refone{fp}{$-1$} in the area $[45^{\circ}N,60^{\circ}N]$, and $0$ elsewhere.

The equations are a subset of those of a more general ocean model \cite[]{niet:07}, where also the longitude dimension is taken into account.
These more general equations are discretized on a longitude-latitude-depth equidistant $n_x \times n_y \times n_z$ grid using a second-order conservative central difference scheme.
An integral condition expressing the overall conservation of salt is also imposed, as the salinity equation is only determined up to an additive constant.
The total number of degrees of freedom is $n = 6 n_x n_y n_z$, as there are six unknowns per point.
The standard spatial resolution used is $n_x = 4, n_y= 32, n_z = 16$ and the solutions of the \cite{niet:07} model are uniform in the zonal direction, with the zonal velocity $u = 0$ and hence satisfy \eqref{GE}.

\begin{figure}[ht]
  \subtop[Bifurcation diagram where a solid line is stable and dashed line is unstable.\label{fig:moc-bifdia-a}]{
    \includegraphics[width=0.48\textwidth,align=c]{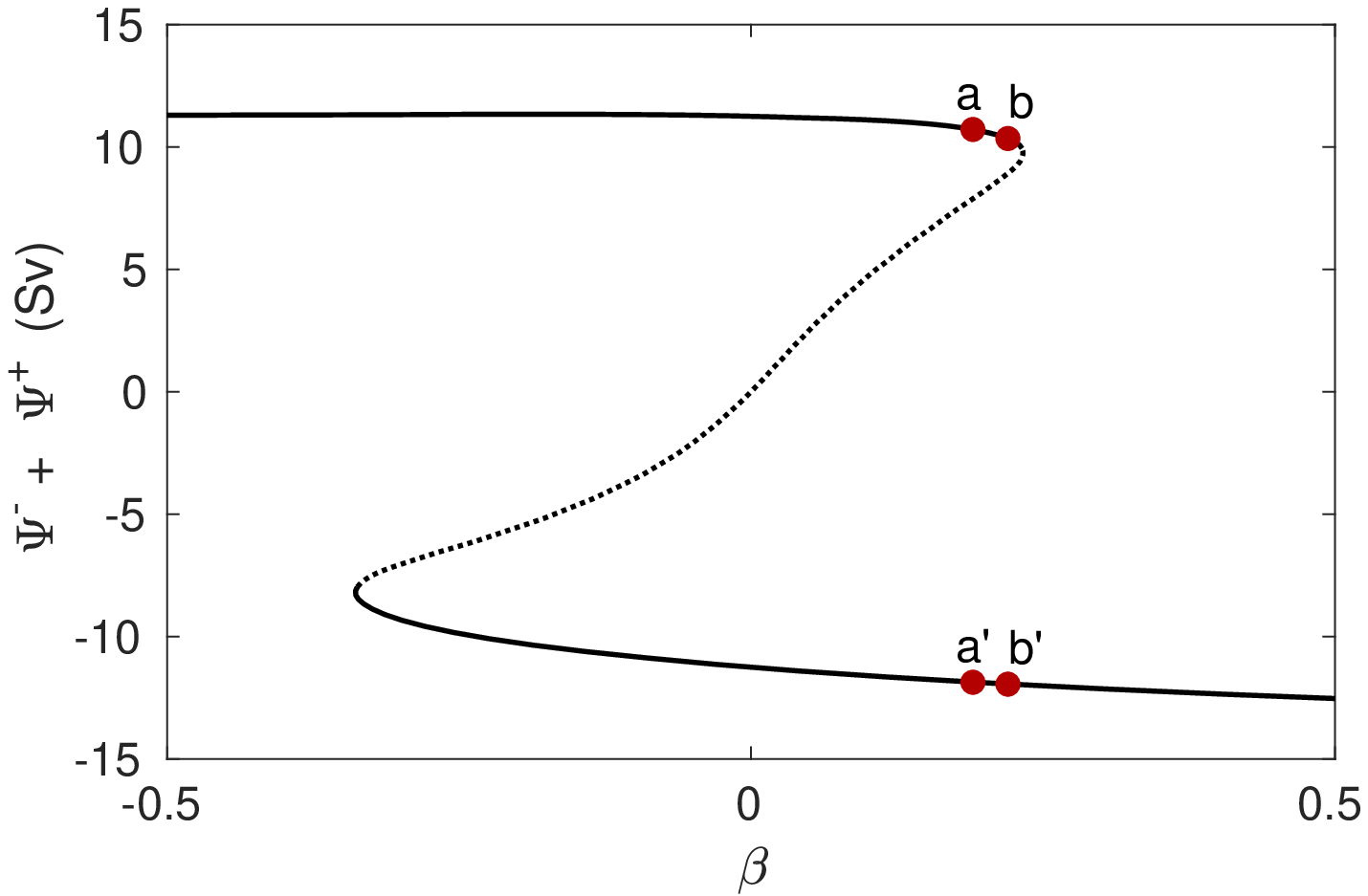}}
  \hspace{0.01\textwidth}
  \subtop[The streamfunction at $a$ and $a'$ respectively.\label{fig:moc-bifdia-b}]{
    \begin{minipage}{0.48\textwidth}
      \centering\includegraphics[width=\textwidth]{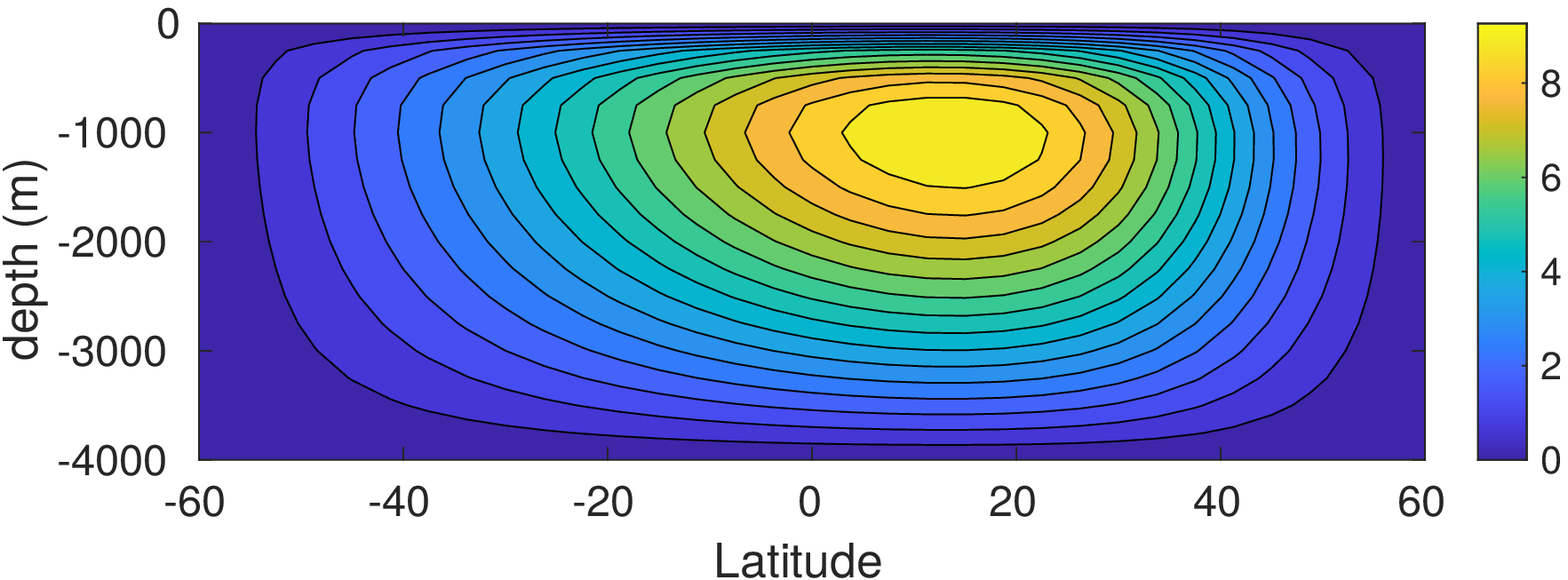}
      \centering\includegraphics[width=\textwidth]{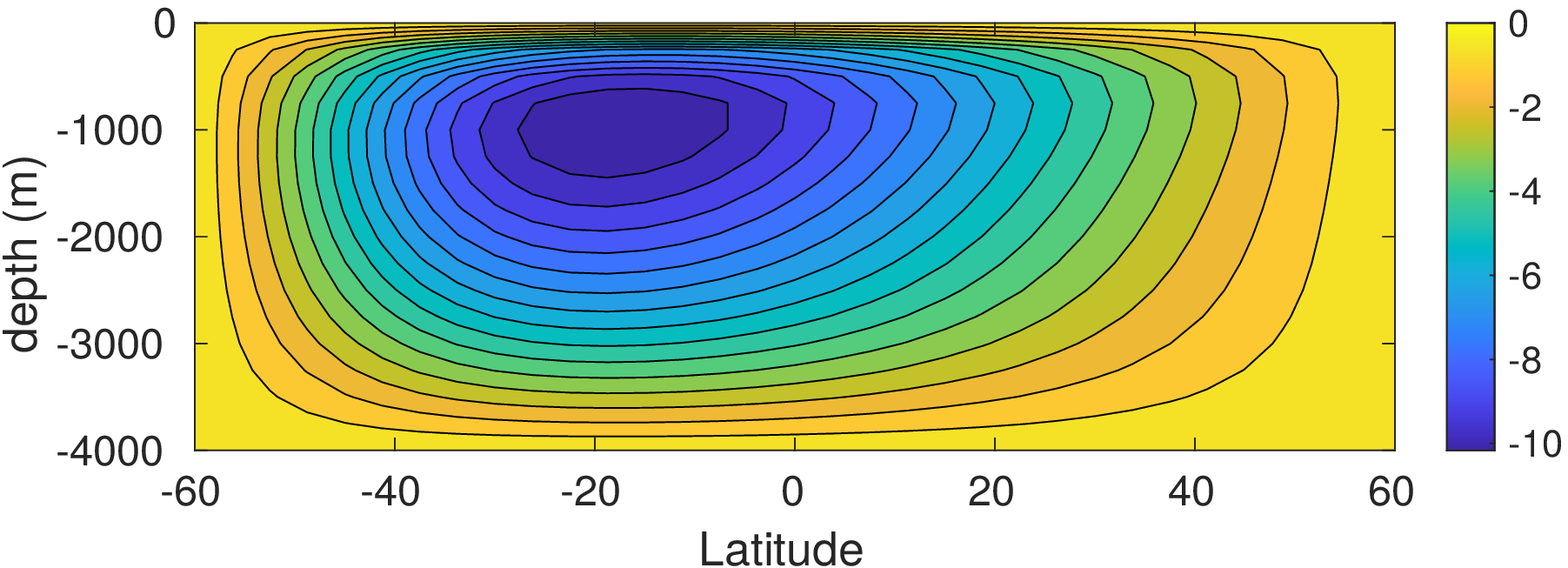}
    \end{minipage}}
  \caption{(a) Bifurcation diagram of the deterministic 2D AMOC model,
    with the forcing as in \eqref{moc-FSd}.
(b) Meridional Overturning Streamfunction patterns at $a$, which represents the current state of the AMOC and $a'$, in which the circulation is reversed, respectively, with contours in Sv.
}
\label{fig:moc-bifdia}
\end{figure}

The bifurcation diagram of the deterministic model for parameters as in \tabref{modelpars} was already calculated in \cite{mheen:13} and is shown in \figref{moc-bifdia-a}.
On the $y$-axis, the sum of the maximum ($\Psi^+$) and minimum ($\Psi^-$) values of the meridional overturning streamfunction $\Psi$ is plotted, where $\Psi$ is defined through
\begin{equation}
  \pd{\Psi}{z} = v  \cos\theta, ~ -  \frac{1}{r_0\cos\theta} \pd{\Psi}{\theta} = w .
\end{equation}
For the calculation of the volume transports (in Sv, where 1 Sv = $10^6$ m$^3$s$^{-1}$), the basin is assumed to have a zonal width of $64^\circ$.
The value of $\Psi^+ + \Psi^-$ is zero when the AMOC is symmetric with respect to the equator.

We are interested in transitions from the top branch to the bottom branch of the bifurcation diagram.
We will investigate transitions at parameter values between $\beta_a=0.190$ and $\beta_b=0.220$.
The stochastic transitions will be from points between $a$ and $b$ to points between $a'$ and $b'$ with the same parameter values.
The pattern of the asymmetric meridional overturning streamfunction at location $a$ with sinking in the northern part of the basin, \refone{moc2}{which represents the current state of the AMOC \cite[]{dijkstra:05},} is shown in \figref{moc-bifdia-b}, as well as the meridional overturning streamfunction at location $a'$ with sinking in the southern part of the basin.
The saddle-node bifurcations are located at $\beta = 0.2320$ and $\beta = -0.3371$.

\subsection{Stochastic freshwater forcing}
For the stochastic model, the freshwater forcing is chosen as
\begin{equation}\label{eq:moc-FS}
  F_s(\theta, t) = (1 + \sigma\zeta(\theta, t))~ \mu F_0 \frac{\cos(\pi\theta/\theta_n)}{\cos(\theta)} + \beta F_p(\theta),
\end{equation}
where $\zeta(\theta, t)$ represents zero-mean white noise with a unit standard deviation, i.e., with $\E[\zeta(\theta, t)] = 0$, $\E[\zeta(\theta, t) \zeta(\theta, s)] = \delta(\theta, t-s)$, $\E[\zeta(\theta, t) \zeta(\eta, t)] = \delta(\theta-\eta, t)$ \refthree{delta}{and $\delta(\theta, t)$ being the Dirac delta function.}
The constant $\sigma$ is the standard deviation of the noise which we set to $\sigma=0.1$.
The noise is additive and is only active in the freshwater component, only present at the surface, meridionally uncorrelated, and has magnitude $\sigma$ of $10\%$ of the background freshwater forcing amplitude at each latitude $\theta$ (see \eqref{moc-FS}).

\subsection{Reaction coordinate}
Since the pressure $p$ is determined up to a constant, and therefore does not have a unique solution, computing the reaction coordinate by using the norm of the state vector as was done in \eqref{react} is not possible.
We instead want to define the reaction coordinate in such a way that the distance between the meridional streamfunctions goes to zero, and therefore we decided to define a norm on the $v$-part of the state.
This gives us a reaction coordinate of the form
\begin{align}
  \phi(\bvec{x}) = \eta - \eta \Exp{-8 d_A^2} + (1-\eta) \Exp{-8 d_B^2},
\end{align}
where, if $\Vert\cdot\Vert_v$ denotes the 2-norm on only the $v$-velocities, $d_A=\Vert \bvec{x} - \sbvec{x}_A\Vert_v / \Vert \sbvec{x}_A - \sbvec{x}_B\Vert_v$ is the normalized difference between $\bvec{x}$ and $\sbvec{x}_A$, and $d_B=\Vert \bvec{x} - \sbvec{x}_B\Vert_v / \Vert \sbvec{x}_A - \sbvec{x}_B\Vert_v$ is the normalized difference between $\bvec{x}$ and $\sbvec{x}_B$, and $\eta=\Vert \sbvec{x}_C - \sbvec{x}_A\Vert_v / \Vert \sbvec{x}_B - \sbvec{x}_A\Vert_v$ is the normalized difference between the unstable steady state $\sbvec{x}_C$ and the stable steady state $\sbvec{x}_A$.
We assume that a trajectory has reached $B$ if $\phi(\bvec{x}) > z_{\max} = 0.75$, since at that point circulation has reversed and we are closer to $\sbvec{x}_B$ as we are to the unstable steady state $\sbvec{x}_C$, as can be seen in \figref{moc-bifdia-a}.

\subsection{Results}
In this section we report on the transition probabilities and compare the results from TAMS with the results from its projected variant.
All optimizations as discussed in \secref{optim} were applied, with exception \refone{checkpointing2}{of the checkpointing strategy and} the parallel computation of trajectories.
The ocean model itself, however, does use a parallel domain decomposition.
Computations are performed on Peregrine, the HPC cluster of the University of Groningen.
Peregrine has nodes with 2 Intel Xeon E5 2680v3 CPUs (24 cores at 2.5GHz) and each node has 128 GB of memory.
For every instance of TAMS, four cores were used to allow for efficient parallelization.
Due to the problem size and the way the convective adjustment is implemented, it did not make sense to use more than four cores per instance.

For our numerical experiments, we investigated the probability of a transition to a reversed circulation within 2000 years and ran TAMS 100 times with 300 trajectories.
The stochastic theta method \eqref{stm} was used with theta equals 0.5 and a time step of 2 years.
\refone{theta2}{Note that a (semi-)implicit method is required due to both the algebraic constraints and the large time step.}
The transition probability, relative error and compensated variance are shown in \figref{results}.

\setlength{\figureWidth}{0.25\textwidth}
\setlength{\figureHeight}{0.25\textwidth}
\pgfplotsset{every axis/.append style={font=\scriptsize}}

\begin{figure}[!ht]
  \subtop[\label{fig:results-a}]{
%
%
\definecolor{mycolor1}{rgb}{0.00000,0.44700,0.74100}%
\begin{tikzpicture}

\begin{axis}[%
width=0.951\figureWidth,
height=\figureHeight,
at={(0\figureWidth,0\figureHeight)},
scale only axis,
xmin=0.19,
xmax=0.22,
xtick={0.19, 0.2, 0.21, 0.22},
xlabel={$\beta$},
ymode=log,
ymin=1e-14,
ymax=1,
yminorticks=true,
ylabel={Transition probability ($\mu_\alpha$)},
axis background/.style={fill=white},
ylabel near ticks,
xlabel near ticks,
axis x line*=bottom,
axis y line*=left
]

\addplot[area legend, draw=none, fill=mycolor1, fill opacity=0.4, forget plot]
table[row sep=crcr] {%
x	y\\
0.22	0.1460959631528\\
0.215	0.0336038990056\\
0.21	0.00242524936198\\
0.205	5.1757749368e-05\\
0.2	9.99998627198954e-17\\
0.195	9.99999983775159e-17\\
0.19	9.99999999995654e-17\\
0.19	5.650324163513e-11\\
0.195	8.45982631097e-08\\
0.2	1.171558718353e-05\\
0.205	0.00049727077888\\
0.21	0.0080416600134\\
0.215	0.0597794335452\\
0.22	0.1881536189172\\
}--cycle;
\addplot [color=mycolor1, line width=0.7pt, forget plot]
  table[row sep=crcr]{%
0.22	0.167124791035\\
0.215	0.0466916662754\\
0.21	0.00523345468769\\
0.205	0.000274514264124\\
0.2	4.4859447217e-06\\
0.195	1.74376005764e-08\\
0.19	8.90663669053e-12\\
};
\end{axis}
\end{tikzpicture}%
}
  \subtop[\label{fig:results-b}]{\input{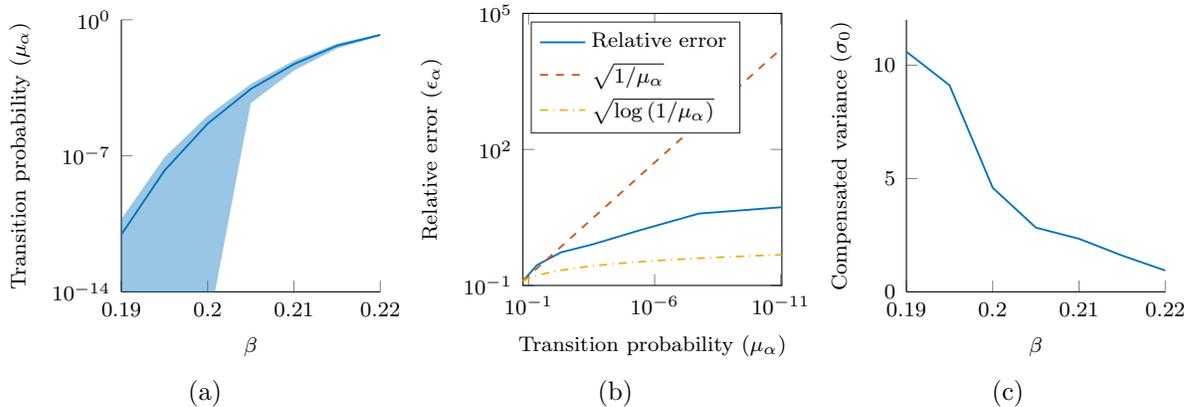}}
  \subtop[\label{fig:results-c}]{
%
%
\definecolor{mycolor1}{rgb}{0.00000,0.44700,0.74100}%
\begin{tikzpicture}

\begin{axis}[%
width=0.951\figureWidth,
height=\figureHeight,
at={(0\figureWidth,0\figureHeight)},
scale only axis,
xmin=0.19,
xmax=0.22,
xtick={0.19, 0.20, 0.21, 0.22},
xlabel={$\beta$},
ymin=0,
ymax=12,
ylabel={Compensated variance ($\sigma_0$)},
axis background/.style={fill=white},
ylabel near ticks,
xlabel near ticks,
axis x line*=bottom,
axis y line*=left
]
\addplot [color=mycolor1, line width=0.7pt, forget plot]
  table[row sep=crcr]{%
0.22	0.940734907663364\\
0.215	1.60128403977301\\
0.21	2.34126065490133\\
0.205	2.8336430228497\\
0.2	4.59254507316909\\
0.195	9.11236585641323\\
0.19	10.5941868794472\\
};
\end{axis}
\end{tikzpicture}%
}
  \caption{Transition probability, relative error and compensated variance on the interval $[\beta_a, \beta_b]$ for $T_{\max}=2000~\text{yr}$ with the shaded area representing the standard deviation.
    To obtain the results we used 100 TAMS experiments with 300 trajectories each.} \label{fig:results}
\end{figure}

We first observe from \figref{results-a} that the transition probability decreases rapidly when moving away from the bifurcation, and that the relative error increases as the transition probability gets smaller.
We also see in \figref{results-b}, however, that the relative error shows much better behavior than the $\sqrt{1/\alpha}$ behavior we would get from a standard Monte Carlo method, as expected.
\refone{var2}{From \figref{results-c} we can conclude that the reaction coordinate can be further improved, since it gets further away from 1 as the transition probability decreases.
The value of the compensated variance remains close to 1, however, in the region of interest where the probability of a transition within the next 2000 years is not negligible.}

To determine how well the projected method performs, we look at accuracy at three different values of $\beta$: $0.190$, $0.205$ and $0.220$, and with increasing sizes of the basis $V$.
The basis is computed by the approximate solution of the generalized Lyapunov equation \eqref{gle} where the relative residual $\Vert R \Vert_2 / \Vert BB^T \Vert_2$ with $R = J(\sbvec{x})VYV^TM^T+MVYV^TJ(\sbvec{x})^T+BB^T$ is decreased to obtain a larger basis.
The results are shown in \figref{results2}.
We see that for all values of $\beta$, the results are very close to the results obtained with the unprojected method when using a basis of rank 400 or larger, at the cost of a slightly larger error.
This is remarkable, because it means that even for small probabilities (i.e.\ $\beta=0.190$), we do not need a larger basis.

\setlength{\figureWidth}{0.25\textwidth}
\setlength{\figureHeight}{0.25\textwidth}
\begin{figure}[!ht]
  \subtop[$\beta=0.190$.\label{fig:results-190}]{
%
%
\definecolor{mycolor1}{rgb}{0.00000,0.44700,0.74100}%
\definecolor{mycolor2}{rgb}{0.85000,0.32500,0.09800}%
\begin{tikzpicture}

\begin{axis}[%
width=0.951\figureWidth,
height=\figureHeight,
at={(0\figureWidth,0\figureHeight)},
scale only axis,
xmin=270,
xmax=614,
xlabel={rank(V)},
ymode=log,
ymin=1e-15,
ymax=1e-08,
yminorticks=true,
ylabel={Transition probability ($\mu_\alpha$)},
axis background/.style={fill=white},
axis x line*=bottom,
axis y line*=left,
ylabel near ticks,
xlabel near ticks,
legend style={at={(0.97,0.03)}, anchor=south east, legend cell align=left, align=left, draw=white!15!black}
]

\addplot[area legend, draw=none, fill=mycolor1, fill opacity=0.4, forget plot]
table[row sep=crcr] {%
x	y\\
1	9.99999999995654e-17\\
101	9.99999999995654e-17\\
201	9.99999999995654e-17\\
301	9.99999999995654e-17\\
401	9.99999999995654e-17\\
501	9.99999999995654e-17\\
601	9.99999999995654e-17\\
701	9.99999999995654e-17\\
801	9.99999999995654e-17\\
901	9.99999999995654e-17\\
1001	9.99999999995654e-17\\
1101	9.99999999995654e-17\\
1201	9.99999999995654e-17\\
1301	9.99999999995654e-17\\
1401	9.99999999995654e-17\\
1401	5.650324163513e-11\\
1301	5.650324163513e-11\\
1201	5.650324163513e-11\\
1101	5.650324163513e-11\\
1001	5.650324163513e-11\\
901	5.650324163513e-11\\
801	5.650324163513e-11\\
701	5.650324163513e-11\\
601	5.650324163513e-11\\
501	5.650324163513e-11\\
401	5.650324163513e-11\\
301	5.650324163513e-11\\
201	5.650324163513e-11\\
101	5.650324163513e-11\\
1	5.650324163513e-11\\
}--cycle;
\addplot [color=mycolor1, line width=0.7pt]
  table[row sep=crcr]{%
1	8.90663669053e-12\\
101	8.90663669053e-12\\
201	8.90663669053e-12\\
301	8.90663669053e-12\\
401	8.90663669053e-12\\
501	8.90663669053e-12\\
601	8.90663669053e-12\\
701	8.90663669053e-12\\
801	8.90663669053e-12\\
901	8.90663669053e-12\\
1001	8.90663669053e-12\\
1101	8.90663669053e-12\\
1201	8.90663669053e-12\\
1301	8.90663669053e-12\\
1401	8.90663669053e-12\\
};
\addlegendentry{TAMS}

\addplot[area legend, draw=none, fill=mycolor2, fill opacity=0.4, forget plot]
table[row sep=crcr] {%
x	y\\
270	1.00000000031877e-16\\
311	1.00000000000001e-16\\
366	1e-16\\
421	9.99999999995654e-17\\
476	1.00000000000373e-16\\
526	9.99999999995654e-17\\
572	9.99999999995654e-17\\
614	9.99999999999693e-17\\
614	1.69117947669e-11\\
572	1.345056383341e-10\\
526	6.265213034143e-11\\
476	2.938634660182e-11\\
421	1.632745713387e-10\\
366	3.19702008397e-14\\
311	1.816766844515e-13\\
270	1.747102714337e-09\\
}--cycle;
\addplot [color=mycolor2, line width=0.7pt]
  table[row sep=crcr]{%
270	6.59967335967e-10\\
311	1.87714526745e-14\\
366	3.360171719e-15\\
421	2.71475740347e-11\\
476	4.99942988542e-12\\
526	9.41868884953e-12\\
572	2.10303990441e-11\\
614	3.375798786e-12\\
};
\addlegendentry{Projected TAMS}

\end{axis}
\end{tikzpicture}%
}
  \subtop[$\beta=0.205$.\label{fig:results-205}]{
%
%
\definecolor{mycolor1}{rgb}{0.00000,0.44700,0.74100}%
\definecolor{mycolor2}{rgb}{0.85000,0.32500,0.09800}%
\begin{tikzpicture}

\begin{axis}[%
width=0.951\figureWidth,
height=\figureHeight,
at={(0\figureWidth,0\figureHeight)},
scale only axis,
xmin=210,
xmax=601,
xlabel={rank(V)},
ymode=log,
ymin=1e-07,
ymax=0.1,
yminorticks=true,
axis background/.style={fill=white},
axis x line*=bottom,
axis y line*=left,
ylabel near ticks,
xlabel near ticks,
legend style={at={(0.97,0.03)}, anchor=south east, legend cell align=left, align=left, draw=white!15!black}
]

\addplot[area legend, draw=none, fill=mycolor1, fill opacity=0.4, forget plot]
table[row sep=crcr] {%
x	y\\
1	5.1757749368e-05\\
101	5.1757749368e-05\\
201	5.1757749368e-05\\
301	5.1757749368e-05\\
401	5.1757749368e-05\\
501	5.1757749368e-05\\
601	5.1757749368e-05\\
701	5.1757749368e-05\\
801	5.1757749368e-05\\
901	5.1757749368e-05\\
1001	5.1757749368e-05\\
1101	5.1757749368e-05\\
1201	5.1757749368e-05\\
1301	5.1757749368e-05\\
1401	5.1757749368e-05\\
1401	0.00049727077888\\
1301	0.00049727077888\\
1201	0.00049727077888\\
1101	0.00049727077888\\
1001	0.00049727077888\\
901	0.00049727077888\\
801	0.00049727077888\\
701	0.00049727077888\\
601	0.00049727077888\\
501	0.00049727077888\\
401	0.00049727077888\\
301	0.00049727077888\\
201	0.00049727077888\\
101	0.00049727077888\\
1	0.00049727077888\\
}--cycle;
\addplot [color=mycolor1, line width=0.7pt]
  table[row sep=crcr]{%
1	0.000274514264124\\
101	0.000274514264124\\
201	0.000274514264124\\
301	0.000274514264124\\
401	0.000274514264124\\
501	0.000274514264124\\
601	0.000274514264124\\
701	0.000274514264124\\
801	0.000274514264124\\
901	0.000274514264124\\
1001	0.000274514264124\\
1101	0.000274514264124\\
1201	0.000274514264124\\
1301	0.000274514264124\\
1401	0.000274514264124\\
};
\addlegendentry{TAMS}

\addplot[area legend, draw=none, fill=mycolor2, fill opacity=0.4, forget plot]
table[row sep=crcr] {%
x	y\\
210	0.0871183552008\\
257	0.00579036124\\
301	1.00000032126485e-15\\
354	1.00000032126485e-15\\
407	9.99986768737693e-16\\
467	4.428941234e-05\\
513	2.14445134499999e-06\\
559	2.366617642e-05\\
601	2.2746649141e-05\\
601	0.000577599206595\\
559	0.000497046309266\\
513	0.000653235585165\\
467	0.00062447670729\\
407	0.000390849452074\\
354	4.02521904454e-05\\
301	0.0001121937630206\\
257	0.01211643001154\\
210	0.1128205260158\\
}--cycle;
\addplot [color=mycolor2, line width=0.7pt]
  table[row sep=crcr]{%
210	0.0999694406083\\
257	0.00895339562577\\
301	5.12030328827e-05\\
354	1.62943449372e-05\\
407	0.000194136161067\\
467	0.000334383059815\\
513	0.000327690018255\\
559	0.000260356242843\\
601	0.000300172927868\\
};
\addlegendentry{Projected TAMS}

\end{axis}
\end{tikzpicture}%
}
  \subtop[$\beta=0.220$.\label{fig:results-220}]{
%
%
\definecolor{mycolor1}{rgb}{0.00000,0.44700,0.74100}%
\definecolor{mycolor2}{rgb}{0.85000,0.32500,0.09800}%
\begin{tikzpicture}

\begin{axis}[%
width=0.951\figureWidth,
height=\figureHeight,
at={(0\figureWidth,0\figureHeight)},
scale only axis,
xmin=252,
xmax=608,
xlabel={rank(V)},
ymode=log,
ymin=0.05,
ymax=0.4,
yminorticks=true,
axis background/.style={fill=white},
axis x line*=bottom,
axis y line*=left,
ylabel near ticks,
xlabel near ticks,
legend style={at={(0.97,0.03)}, anchor=south east, legend cell align=left, align=left, draw=white!15!black}
]

\addplot[area legend, draw=none, fill=mycolor1, fill opacity=0.4, forget plot]
table[row sep=crcr] {%
x	y\\
1	0.1460959631528\\
101	0.1460959631528\\
201	0.1460959631528\\
301	0.1460959631528\\
401	0.1460959631528\\
501	0.1460959631528\\
601	0.1460959631528\\
701	0.1460959631528\\
801	0.1460959631528\\
901	0.1460959631528\\
1001	0.1460959631528\\
1101	0.1460959631528\\
1201	0.1460959631528\\
1301	0.1460959631528\\
1401	0.1460959631528\\
1401	0.1881536189172\\
1301	0.1881536189172\\
1201	0.1881536189172\\
1101	0.1881536189172\\
1001	0.1881536189172\\
901	0.1881536189172\\
801	0.1881536189172\\
701	0.1881536189172\\
601	0.1881536189172\\
501	0.1881536189172\\
401	0.1881536189172\\
301	0.1881536189172\\
201	0.1881536189172\\
101	0.1881536189172\\
1	0.1881536189172\\
}--cycle;
\addplot [color=mycolor1, line width=0.7pt]
  table[row sep=crcr]{%
1	0.167124791035\\
101	0.167124791035\\
201	0.167124791035\\
301	0.167124791035\\
401	0.167124791035\\
501	0.167124791035\\
601	0.167124791035\\
701	0.167124791035\\
801	0.167124791035\\
901	0.167124791035\\
1001	0.167124791035\\
1101	0.167124791035\\
1201	0.167124791035\\
1301	0.167124791035\\
1401	0.167124791035\\
};
\addlegendentry{TAMS}

\addplot[area legend, draw=none, fill=mycolor2, fill opacity=0.4, forget plot]
table[row sep=crcr] {%
x	y\\
252	0.2751905946303\\
297	0.1744748676103\\
351	0.1437777633384\\
405	0.1349557092375\\
466	0.1304617311035\\
519	0.1400262022026\\
559	0.1400271777498\\
608	0.1350558611464\\
608	0.1840428542536\\
559	0.1888443036222\\
519	0.1905987409854\\
466	0.1778644728245\\
405	0.1789436059465\\
351	0.1984109843196\\
297	0.2270447032537\\
252	0.3321839400377\\
}--cycle;
\addplot [color=mycolor2, line width=0.7pt]
  table[row sep=crcr]{%
252	0.303687267334\\
297	0.200759785432\\
351	0.171094373829\\
405	0.156949657592\\
466	0.154163101964\\
519	0.165312471594\\
559	0.164435740686\\
608	0.1595493577\\
};
\addlegendentry{Projected TAMS}

\end{axis}
\end{tikzpicture}%
}
  \caption{Transition probability at different values of $\beta$ and with increasing size of the basis $V$.
    The shaded area represents the standard deviation and $T_{\max}=2000~\text{yr}$.
    To obtain the results we used 100 TAMS experiments with 300 trajectories each.} \label{fig:results2}
\end{figure}
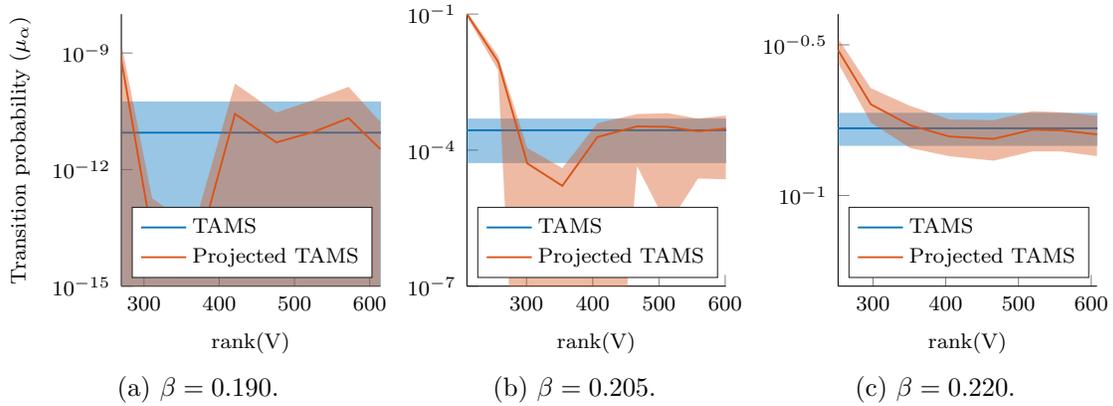

\begin{figure}[!ht]
  \centering
  \input{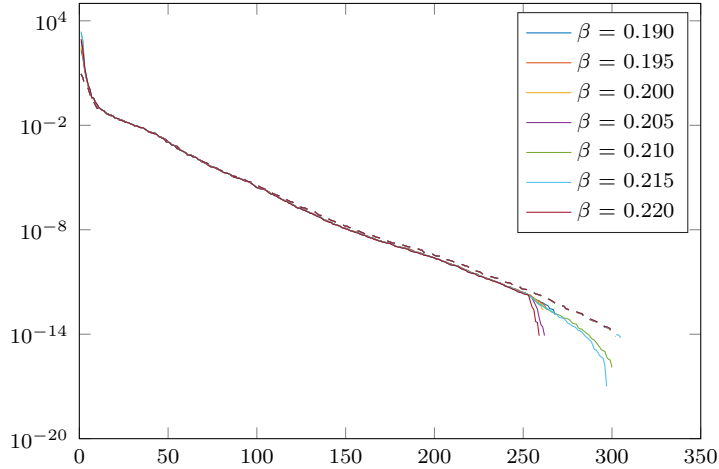}
  \caption{Leading eigenvalues of the covariance matrix at the top (solid) and bottom (dashed) branch as determined by the Lyapunov solver RAILS \cite[]{baars:17} with a relative residual tolerance of $10^{-10}$.}
\label{fig:evs}
\end{figure}

\llabel{acommon}
A common a priori method of determining how large the basis should be is by looking at the leading eigenvalues of the corresponding matrices.
In our case these are the covariance matrices at two branches.
As can be seen in \figref{evs}, we never observe a sudden decrease of the eigenvalues, which would indicate that a sufficient amount has been determined.
Indeed, we found that we are able to use the same basis size for all probabilities that we compute.
This motivates to investigate whether, for a larger application, we would be able to determine the proper basis size once by looking at a figure such as \figref{results-220}, and then use that basis size for all further experiments.
However, this is outside the scope of this paper.
Note that computing \figref{results-220} is still relatively cheap when the probability is large.

\setlength{\figureWidth}{0.6\textwidth}
\setlength{\figureHeight}{0.4\textwidth}

In \tabref{cost} we show the computational cost of the unprojected method compared to the projected method for different ranks of the basis $V$.
The values we show were determined at $\beta=0.220$, but these are actually independent of the value of $\beta$.
Here we observe that for a basis of rank $~400$, which we deemed sufficient, the projected method is twice as fast and uses 97\% less memory.

\begin{table}[ht]
  \centering
  \begin{tabular}{l|lllll}
    $\textrm{rank}(V)$ & $t(s)$ & Mem (kB)\\\hline
      $-$ &     $0.32$ &   96.00 kB\\\hline
    $252$ &     $0.12$ &    1.97 kB\\
    $297$ &     $0.13$ &    2.32 kB\\
    $351$ &     $0.14$ &    2.74 kB\\
    $405$ &     $0.17$ &    3.16 kB\\
    $466$ &     $0.19$ &    3.64 kB\\
    $519$ &     $0.22$ &    4.05 kB\\
    $559$ &     $0.24$ &    4.37 kB\\
    $608$ &     $0.26$ &    4.75 kB\\
  \end{tabular}
  \caption{
    Computational time per time step and memory cost per state vector of the unprojected method (first row) and the projected method at increasing sizes of $V$.
    \label{tab:cost}}
\end{table}
\section{Summary and Discussion}

The Adaptive Multilevel Splitting (AMS) method and its Trajectory variant (TAMS) are very promising numerical techniques to study rare events and in particular transitions between different equilibria of stochastic dynamical systems.
However, in their original form they cannot be applied to high-dimensional dynamical systems, i.e.\ as derived from three-dimensional ocean models.
Therefore, we implemented a projected time-stepping method, which can be applied to any of the methods in the Generalized Adaptive Multilevel Splitting framework \cite[]{brehier:16}.
This approach solves the problem in a reduced space, which is obtained from a Galerkin-like projection with the low-rank solution of a generalized Lyapunov equation.
The results obtained with the projected time-stepping approach are very similar to the ones obtained with standard TAMS.
The main benefit of the projected time-stepping method is the reduced computational cost, especially in terms of memory usage, where only $3\%$ of the amount of memory that is required for standard TAMS is used.
\refone{checkpointing3}{Since memory is the limiting factor in many applications, this reduction, along with all other optimizations such as the checkpointing strategy we mentioned, will allow a larger application potential, without necessarily negatively impacting the computational time.}

The benefit in terms of memory usage is present even if an explicit time stepping method is used.
Downsides of using the projection in combination with an explicit method, however, are a higher computational cost due to the projection that has to be performed in every time step and the requirement of having access to the Jacobian matrix to be able to compute the low-rank approximation of the covariance matrix.
In case of implicit time stepping, however, the Jacobian matrix is usually already available, and solving the projected system is often cheaper than solving with the full Jacobian matrix.

As an application, we computed transition probabilities for a quasi two-dimensional model of the Atlantic Meridional Overturning Circulation (AMOC).
The model consists of a system of stochastic partial differential equations, representing the equations for mass, momentum, temperature and salinity in the oceanic basin, resulting (when discretized) in a dynamical system with a total of 12288 degrees of freedom.
While this is a highly idealized model, it captures the essential feedback (the salt-advection feedback) which is responsible for the sensitivity of the AMOC to freshwater perturbations.
Although we investigated only a few cases, the transition probability of the AMOC in the two-dimensional model decreases drastically when moving away from the saddle-node bifurcation.
This indicates that, in this model, a stochastic transition is not very likely to occur unless the AMOC is close to critical conditions.

\reftwo{3d}{It would be interesting to see if the methodology can be extended to three-dimensional ocean models for which one can calculate bifurcation diagrams.
In \cite{baars:17}, we found for a primitive three-dimensional example where the noise is fully uncorrelated (worst case), that the size of the space that comes from the generalized Lyapunov equation increases with $\mathcal{O}(n)$.
Therefore it is possible that for three-dimensional ocean models the cost of solving with the projected Jacobian matrix becomes dominant, and eventually more expensive than computations with the original Jacobian matrix, but the memory cost will still be at most $3\%$ of the original TAMS.
Whether we can suffice with predicting the required size of the basis from simulations with a high transition probability is still an open question.}

\refone{noise}{Additionally, more sophisticated noise models may be used, such as correlated additive and multiplicative (CAM) noise, which is commonly used in climate science.
The methodology from this paper can also be used for multiplicative types of noise, but in that case the generalized Lyapunov equations have to be extended with a bilinear term.
We refer to the resulting equations as extended generalized Lyapunov equations.
These may be solved by combining the methodology from \cite{shank:15} and \cite{baars:17}.}

\refone{var3}{Something that still needs further investigation is the reaction coordinate.
In this paper, we just chose a reaction coordinate that works reasonably well for the AMOC application, but the fact that the compensated variance moves away from 1 as the transition probability becomes smaller suggests that it is possible to find more efficient reaction coordinates.
It has also been shown theoretically for systems with a few degrees of freedom that paths that go through $\sbvec{x}_C$ should be favored as they more closely represent the committor \cite[]{metzner:06}.
Further incorporating terms of the form $\Vert x - x_C \Vert$ may help with this \cite[]{rolland:15, rolland:15b}.
It would be \refone{favorable}{favorable} if it were possible to find a reaction coordinate that works well for a wide variety of high-dimensional dynamical systems, and work on this is in progress.}

\subsection*{Acknowledgments}
The authors would like to thank Jeroen Wouters and Daan Crommelin for the useful discussions.
We would also like to thank the reviewers for their very helpful and constructive comments, which greatly improved the paper.
This work is part of the Mathematics of Planet Earth research program with project number 657.014.007, which is financed by the Netherlands Organization for Scientific Research (NWO) (SB and FW), the SMCM project of the Netherlands eScience Center (NLeSC) with project number 027.017.G02 (SB, FW and HD) and the European Union's Horizon 2020 research and innovation program for the ITN CRITICS under Grant Agreement Number 643073 (DC and HD).

\bibliographystyle{abbrvnatdoi}
\bibliography{TAMS}

\begin{thebibliography}{47}
\providecommand{\natexlab}[1]{#1}
\providecommand{\url}[1]{\texttt{#1}}
\expandafter\ifx\csname urlstyle\endcsname\relax
  \providecommand{\doi}[1]{doi: #1}\else
  \providecommand{\doi}[1]{doi: \urlstyle{same}
  \href{https://dx.doi.org/#1}{\nolinkurl{#1}}}\fi

\bibitem[Baars(2019)]{baars:19}
S.~Baars.
\newblock \emph{Numerical methods for studying transition probabilities in
  stochastic ocean-climate models}.
\newblock PhD thesis, University of Groningen, Jun 2019.

\bibitem[Baars et~al.(2017)Baars, Viebahn, Mulder, Kuehn, Wubs, and
  Dijkstra]{baars:17}
S.~Baars, J.~P. Viebahn, T.~E. Mulder, C.~Kuehn, F.~W. Wubs, and H.~A.
  Dijkstra.
\newblock Continuation of probability density functions using a generalized
  {L}yapunov approach.
\newblock \emph{Journal of Computational Physics}, 336:\penalty0 627--643, May
  2017.
\newblock ISSN 0021-9991.
\newblock \doi{10.1016/j.jcp.2017.02.021}.

\bibitem[Bouchet and Reygner(2016)]{bouchet:16}
F.~Bouchet and J.~Reygner.
\newblock Generalisation of the {E}yring--{K}ramers transition rate formula to
  irreversible diffusion processes.
\newblock \emph{Annales Henri Poincar{\'e}}, 17\penalty0 (12):\penalty0
  3499--3532, Jun 2016.
\newblock ISSN 1424-0661.
\newblock \doi{10.1007/s00023-016-0507-4}.

\bibitem[Bouchet et~al.(2014)Bouchet, Laurie, and Zaboronski]{bouchet:14}
F.~Bouchet, J.~Laurie, and O.~Zaboronski.
\newblock Langevin dynamics, large deviations and instantons for the
  quasi-geostrophic model and two-dimensional {E}uler equations.
\newblock \emph{Journal of Statistical Physics}, 156\penalty0 (6):\penalty0
  1066--1092, Jul 2014.
\newblock ISSN 1572-9613.
\newblock \doi{10.1007/s10955-014-1052-5}.

\bibitem[Bouchet et~al.(2019)Bouchet, Rolland, and Simonnet]{bouchet:19}
F.~Bouchet, J.~Rolland, and E.~Simonnet.
\newblock Rare event algorithm links transitions in turbulent flows with
  activated nucleations.
\newblock \emph{Physical Review Letters}, 122\penalty0 (7), Feb 2019.
\newblock ISSN 1079-7114.
\newblock \doi{10.1103/physrevlett.122.074502}.

\bibitem[Br{\'e}hier et~al.(2016)Br{\'e}hier, Gazeau, Gouden{\`e}ge,
  Leli{\`e}vre, and Rousset]{brehier:16}
C.-E. Br{\'e}hier, M.~Gazeau, L.~Gouden{\`e}ge, T.~Leli{\`e}vre, and
  M.~Rousset.
\newblock Unbiasedness of some generalized adaptive multilevel splitting
  algorithms.
\newblock \emph{The Annals of Applied Probability}, 26\penalty0 (6):\penalty0
  3559--3601, Dec 2016.
\newblock ISSN 1050-5164.
\newblock \doi{10.1214/16-aap1185}.

\bibitem[Castellana et~al.(2019)Castellana, Baars, Wubs, and
  Dijkstra]{castellana:19}
D.~Castellana, S.~Baars, F.~W. Wubs, and H.~A. Dijkstra.
\newblock Transition probabilities of noise-induced transitions of the atlantic
  ocean circulation.
\newblock \emph{Scientific Reports}, 9\penalty0 (1), Dec 2019.
\newblock ISSN 2045-2322.
\newblock \doi{10.1038/s41598-019-56435-6}.

\bibitem[Cazemier et~al.(1998)Cazemier, Verstappen, and Veldman]{cazemier:98}
W.~Cazemier, R.~W. C.~P. Verstappen, and A.~E.~P. Veldman.
\newblock Proper orthogonal decomposition and low-dimensional models for driven
  cavity flows.
\newblock \emph{Physics of Fluids}, 10\penalty0 (7):\penalty0 1685--1699, Jul
  1998.
\newblock ISSN 1089-7666.
\newblock \doi{10.1063/1.869686}.

\bibitem[C{\'e}rou and Guyader(2007)]{cerou:07}
F.~C{\'e}rou and A.~Guyader.
\newblock Adaptive multilevel splitting for rare event analysis.
\newblock \emph{Stochastic Analysis and Applications}, 25\penalty0
  (2):\penalty0 417--443, Feb 2007.
\newblock ISSN 1532-9356.
\newblock \doi{10.1080/07362990601139628}.

\bibitem[C{\'e}rou et~al.(2011)C{\'e}rou, Guyader, Leli{\`e}vre, and
  Pommier]{cerou:11}
F.~C{\'e}rou, A.~Guyader, T.~Leli{\`e}vre, and D.~Pommier.
\newblock A multiple replica approach to simulate reactive trajectories.
\newblock \emph{The Journal of Chemical Physics}, 134\penalty0 (5):\penalty0
  054108, Feb 2011.
\newblock ISSN 1089-7690.
\newblock \doi{10.1063/1.3518708}.

\bibitem[de~Niet et~al.(2007)de~Niet, Wubs, van Scheltinga, and
  Dijkstra]{niet:07}
A.~de~Niet, F.~Wubs, A.~T. van Scheltinga, and H.~A. Dijkstra.
\newblock A tailored solver for bifurcation analysis of ocean-climate models.
\newblock \emph{Journal of Computational Physics}, 227\penalty0 (1):\penalty0
  654--679, Nov 2007.
\newblock ISSN 0021-9991.
\newblock \doi{10.1016/j.jcp.2007.08.006}.

\bibitem[den Toom et~al.(2011)den Toom, Dijkstra, and Wubs]{toom:11}
M.~den Toom, H.~A. Dijkstra, and F.~W. Wubs.
\newblock Spurious multiple equilibria introduced by convective adjustment.
\newblock \emph{Ocean Modelling}, 38\penalty0 (1-2):\penalty0 126--137, Jan
  2011.
\newblock ISSN 1463-5003.
\newblock \doi{10.1016/j.ocemod.2011.02.009}.

\bibitem[Dijkstra(2005)]{dijkstra:05}
H.~A. Dijkstra.
\newblock \emph{Nonlinear Physical Oceanography}.
\newblock Springer Netherlands, 2005.
\newblock ISBN 9781402022623.
\newblock \doi{10.1007/1-4020-2263-8}.

\bibitem[E et~al.(2004)E, Ren, and Vanden-Eijnden]{e:04}
W.~E, W.~Ren, and E.~Vanden-Eijnden.
\newblock Minimum action method for the study of rare events.
\newblock \emph{Communications on Pure and Applied Mathematics}, 57\penalty0
  (5):\penalty0 637--656, 2004.
\newblock ISSN 1097-0312.
\newblock \doi{10.1002/cpa.20005}.

\bibitem[Eyring(1935)]{eyring:35}
H.~Eyring.
\newblock The activated complex in chemical reactions.
\newblock \emph{The Journal of Chemical Physics}, 3\penalty0 (2):\penalty0
  107--115, 1935.
\newblock \doi{10.1063/1.1749604}.

\bibitem[Freidlin and Wentzell(1984)]{freidlin:84}
M.~I. Freidlin and A.~D. Wentzell.
\newblock \emph{Random Perturbations}.
\newblock Springer US, 1984.
\newblock ISBN 9781468401769.
\newblock \doi{10.1007/978-1-4684-0176-9_2}.

\bibitem[Gardiner(1985)]{gardiner:85}
C.~W. Gardiner.
\newblock \emph{Handbook of Stochastic Methods for Physics, Chemistry and the
  Natural Sciences}.
\newblock Springer Berlin Heidelberg, 1985.
\newblock ISBN 9783662024522.
\newblock \doi{10.1007/978-3-662-02452-2}.

\bibitem[Grafke et~al.(2017)Grafke, Sch{\"a}fer, and Vanden-Eijnden]{grafke:17}
T.~Grafke, T.~Sch{\"a}fer, and E.~Vanden-Eijnden.
\newblock Long term effects of small random perturbations on dynamical systems:
  Theoretical and computational tools.
\newblock \emph{Recent Progress and Modern Challenges in Applied Mathematics,
  Modeling and Computational Science}, pages 17--55, 2017.
\newblock ISSN 2194-1564.
\newblock \doi{10.1007/978-1-4939-6969-2_2}.

\bibitem[Hartmann(2011)]{hartmann:11}
C.~Hartmann.
\newblock Balanced model reduction of partially observed langevin equations: an
  averaging principle.
\newblock \emph{Mathematical and Computer Modelling of Dynamical Systems},
  17\penalty0 (5):\penalty0 463--490, Oct 2011.
\newblock ISSN 1744-5051.
\newblock \doi{10.1080/13873954.2011.576517}.

\bibitem[Hartmann and Sch{\"u}tte(2005)]{hartmann:05}
C.~Hartmann and C.~Sch{\"u}tte.
\newblock A constrained hybrid monte-carlo algorithm and the problem of
  calculating the free energy in several variables.
\newblock \emph{ZAMM}, 85\penalty0 (10):\penalty0 700--710, Oct 2005.
\newblock ISSN 1521-4001.
\newblock \doi{10.1002/zamm.200410218}.

\bibitem[Hartmann et~al.(2016)Hartmann, Sch{\"u}tte, and Zhang]{hartmann:16}
C.~Hartmann, C.~Sch{\"u}tte, and W.~Zhang.
\newblock Model reduction algorithms for optimal control and importance
  sampling of diffusions.
\newblock \emph{Nonlinearity}, 29\penalty0 (8):\penalty0 2298--2326, Jun 2016.
\newblock ISSN 1361-6544.
\newblock \doi{10.1088/0951-7715/29/8/2298}.

\bibitem[Hern{\'a}ndez et~al.(2018)Hern{\'a}ndez, Wayment-Steele, Sultan,
  Husic, and Pande]{hernandez:18}
C.~X. Hern{\'a}ndez, H.~K. Wayment-Steele, M.~M. Sultan, B.~E. Husic, and V.~S.
  Pande.
\newblock Variational encoding of complex dynamics.
\newblock \emph{Physical Review E}, 97\penalty0 (6), Jun 2018.
\newblock ISSN 2470-0053.
\newblock \doi{10.1103/physreve.97.062412}.

\bibitem[Kahn and Harris(1951)]{kahn:51}
H.~Kahn and T.~E. Harris.
\newblock Estimation of particle transmission by random sampling.
\newblock \emph{National Bureau of Standards applied mathematics series},
  12:\penalty0 27--30, 1951.

\bibitem[Keller(1977)]{keller:77}
H.~B. Keller.
\newblock Numerical solution of bifurcation and nonlinear eigenvalue problems.
\newblock In P.~H. Rabinowitz, editor, \emph{Applications of Bifurcation
  Theory}, pages 359--384. Academic Press, New York, U.S.A., 1977.

\bibitem[Kloeden and Platen(1992)]{kloeden:92}
P.~E. Kloeden and E.~Platen.
\newblock \emph{Numerical Solution of Stochastic Differential Equations}.
\newblock Springer Berlin Heidelberg, 1992.
\newblock ISBN 9783662126165.
\newblock \doi{10.1007/978-3-662-12616-5}.

\bibitem[Kramers(1940)]{kramers:40}
H.~A. Kramers.
\newblock Brownian motion in a field of force and the diffusion model of
  chemical reactions.
\newblock \emph{Physica}, 7\penalty0 (4):\penalty0 284--304, Apr 1940.
\newblock ISSN 0031-8914.
\newblock \doi{10.1016/s0031-8914(40)90098-2}.

\bibitem[Laurie and Bouchet(2015)]{laurie:15}
J.~Laurie and F.~Bouchet.
\newblock Computation of rare transitions in the barotropic quasi-geostrophic
  equations.
\newblock \emph{New Journal of Physics}, 17\penalty0 (1):\penalty0 015009, Jan
  2015.
\newblock ISSN 1367-2630.
\newblock \doi{10.1088/1367-2630/17/1/015009}.

\bibitem[Lestang et~al.(2018)Lestang, Ragone, Br{\'e}hier, Herbert, and
  Bouchet]{lestang:18}
T.~Lestang, F.~Ragone, C.-E. Br{\'e}hier, C.~Herbert, and F.~Bouchet.
\newblock Computing return times or return periods with rare event algorithms.
\newblock \emph{Journal of Statistical Mechanics: Theory and Experiment},
  2018\penalty0 (4):\penalty0 043213, Apr 2018.
\newblock ISSN 1742-5468.
\newblock \doi{10.1088/1742-5468/aab856}.

\bibitem[Lo\`eve(1955)]{loeve:55}
M.~Lo\`eve.
\newblock \emph{Probability Theory: Foundations, Random Sequences}.
\newblock van Nostrand, 1955.

\bibitem[Lumley(1967)]{lumley:67}
J.~L. Lumley.
\newblock The structure of inhomogeneous turbulent flows.
\newblock \emph{Atmospheric turbulence and wave propagation}, pages 166--178,
  1967.

\bibitem[Metzner et~al.(2006)Metzner, Sch{\"u}tte, and
  Vanden-Eijnden]{metzner:06}
P.~Metzner, C.~Sch{\"u}tte, and E.~Vanden-Eijnden.
\newblock Illustration of transition path theory on a collection of simple
  examples.
\newblock \emph{The Journal of Chemical Physics}, 125\penalty0 (8):\penalty0
  084110, Aug 2006.
\newblock ISSN 1089-7690.
\newblock \doi{10.1063/1.2335447}.

\bibitem[Mohamad et~al.(2016)Mohamad, Cousins, and Sapsis]{mohamad:16}
M.~A. Mohamad, W.~Cousins, and T.~P. Sapsis.
\newblock A probabilistic decomposition-synthesis method for the quantification
  of rare events due to internal instabilities.
\newblock \emph{Journal of Computational Physics}, 322:\penalty0 288--308, Oct
  2016.
\newblock ISSN 0021-9991.
\newblock \doi{10.1016/j.jcp.2016.06.047}.

\bibitem[Moral(2013)]{moral:13}
P.~D. Moral.
\newblock \emph{Mean Field Simulation for {M}onte {C}arlo Integration}.
\newblock Chapman and Hall/CRC, May 2013.
\newblock ISBN 9781466504172.
\newblock \doi{10.1201/b14924}.

\bibitem[Moral and Garnier(2005)]{moral:05}
P.~D. Moral and J.~Garnier.
\newblock Genealogical particle analysis of rare events.
\newblock \emph{The Annals of Applied Probability}, 15\penalty0 (4):\penalty0
  2496--2534, Nov 2005.
\newblock ISSN 1050-5164.
\newblock \doi{10.1214/105051605000000566}.

\bibitem[Mukhin et~al.(2015)Mukhin, Kondrashov, Loskutov, Gavrilov, Feigin, and
  Ghil]{mukhin:15}
D.~Mukhin, D.~Kondrashov, E.~Loskutov, A.~Gavrilov, A.~Feigin, and M.~Ghil.
\newblock Predicting critical transitions in {ENSO} models. part {II:}
  spatially dependent models.
\newblock \emph{Journal of Climate}, 28\penalty0 (5):\penalty0 1962--1976, Feb
  2015.
\newblock ISSN 1520-0442.
\newblock \doi{10.1175/jcli-d-14-00240.1}.

\bibitem[Rolland and Simonnet(2015)]{rolland:15}
J.~Rolland and E.~Simonnet.
\newblock Statistical behaviour of adaptive multilevel splitting algorithms in
  simple models.
\newblock \emph{Journal of Computational Physics}, 283:\penalty0 541--558, Feb
  2015.
\newblock ISSN 0021-9991.
\newblock \doi{10.1016/j.jcp.2014.12.009}.

\bibitem[Rolland et~al.(2015)Rolland, Bouchet, and Simonnet]{rolland:15b}
J.~Rolland, F.~Bouchet, and E.~Simonnet.
\newblock Computing transition rates for the {1-D} stochastic
  {G}inzburg--{L}andau--{A}llen--{C}ahn equation for finite-amplitude noise
  with a rare event algorithm.
\newblock \emph{Journal of Statistical Physics}, 162\penalty0 (2):\penalty0
  277--311, Nov 2015.
\newblock ISSN 1572-9613.
\newblock \doi{10.1007/s10955-015-1417-4}.

\bibitem[Rosenbluth and Rosenbluth(1955)]{rosenbluth:55}
M.~N. Rosenbluth and A.~W. Rosenbluth.
\newblock {M}onte {C}arlo calculation of the average extension of molecular
  chains.
\newblock \emph{The Journal of Chemical Physics}, 23\penalty0 (2):\penalty0
  356--359, Feb 1955.
\newblock ISSN 1089-7690.
\newblock \doi{10.1063/1.1741967}.

\bibitem[Rubino and Tuffin(2009)]{rubino:09}
G.~Rubino and B.~Tuffin.
\newblock \emph{Rare Event Simulation using {M}onte {C}arlo Methods}.
\newblock John Wiley \& Sons, Ltd, Mar 2009.
\newblock ISBN 9780470772690.
\newblock \doi{10.1002/9780470745403}.

\bibitem[Shank et~al.(2015)Shank, Simoncini, and Szyld]{shank:15}
S.~D. Shank, V.~Simoncini, and D.~B. Szyld.
\newblock Efficient low-rank solution of generalized lyapunov equations.
\newblock \emph{Numerische Mathematik}, 134\penalty0 (2):\penalty0 327--342,
  Nov 2015.
\newblock ISSN 0945-3245.
\newblock \doi{10.1007/s00211-015-0777-7}.

\bibitem[Simonnet(2014)]{simonnet:14}
E.~Simonnet.
\newblock Combinatorial analysis of the adaptive last particle method.
\newblock \emph{Statistics and Computing}, 26\penalty0 (1-2):\penalty0
  211--230, Jul 2014.
\newblock ISSN 1573-1375.
\newblock \doi{10.1007/s11222-014-9489-6}.

\bibitem[Stommel(1961)]{stommel:61}
H.~Stommel.
\newblock Thermohaline convection with two stable regimes of flow.
\newblock \emph{Tellus B}, 13\penalty0 (2), May 1961.
\newblock ISSN 0280-6509.
\newblock \doi{10.3402/tellusb.v13i2.12985}.

\bibitem[van~der Mheen et~al.(2013)van~der Mheen, Dijkstra, Gozolchiani, den
  Toom, Feng, Kurths, and Hernandez-Garcia]{mheen:13}
M.~van~der Mheen, H.~A. Dijkstra, A.~Gozolchiani, M.~den Toom, Q.~Feng,
  J.~Kurths, and E.~Hernandez-Garcia.
\newblock Interaction network based early warning indicators for the {A}tlantic
  {MOC} collapse.
\newblock \emph{Geophysical Research Letters}, 40\penalty0 (11):\penalty0
  2714--2719, Jun 2013.
\newblock ISSN 0094-8276.
\newblock \doi{10.1002/grl.50515}.

\bibitem[Vanden-Eijnden and Heymann(2008)]{vanden-eijnden:08}
E.~Vanden-Eijnden and M.~Heymann.
\newblock The geometric minimum action method for computing minimum energy
  paths.
\newblock \emph{The Journal of Chemical Physics}, 128\penalty0 (6):\penalty0
  061103, Feb 2008.
\newblock ISSN 1089-7690.
\newblock \doi{10.1063/1.2833040}.

\bibitem[Wouters and Bouchet(2016)]{wouters:16}
J.~Wouters and F.~Bouchet.
\newblock Rare event computation in deterministic chaotic systems using
  genealogical particle analysis.
\newblock \emph{Journal of Physics A: Mathematical and Theoretical},
  49\penalty0 (37):\penalty0 374002, Aug 2016.
\newblock ISSN 1751-8121.
\newblock \doi{10.1088/1751-8113/49/37/374002}.

\bibitem[Yvinec et~al.(2012)Yvinec, D'Orsogna, and Chou]{yvinec:12}
R.~Yvinec, M.~R. D'Orsogna, and T.~Chou.
\newblock First passage times in homogeneous nucleation and self-assembly.
\newblock \emph{The Journal of Chemical Physics}, 137\penalty0 (24):\penalty0
  244107, Dec 2012.
\newblock ISSN 1089-7690.
\newblock \doi{10.1063/1.4772598}.

\bibitem[Zhou et~al.(2008)Zhou, Ren, and E]{zhou:08}
X.~Zhou, W.~Ren, and W.~E.
\newblock Adaptive minimum action method for the study of rare events.
\newblock \emph{The Journal of Chemical Physics}, 128\penalty0 (10):\penalty0
  104111, Mar 2008.
\newblock ISSN 1089-7690.
\newblock \doi{10.1063/1.2830717}.

\end{thebibliography}

\end{document}